\documentclass[ reprint,
superscriptaddress,
nobibnotes,
amsmath,amssymb,
aps,
prx,onecolumn
]{revtex4-2}

\usepackage[utf8]{inputenc}
\usepackage{amsmath}
\usepackage{amsfonts}
\usepackage{amssymb}
\usepackage{graphicx}
\usepackage{color}
\usepackage{txfonts}
\usepackage{float}
\usepackage[titletoc]{appendix}
\usepackage[colorlinks={true}]{hyperref}
\hypersetup{
    citecolor={blue},
    filecolor={blue},
    linkcolor={blue},
    urlcolor={black}
}

\usepackage{graphicx,epsfig,epstopdf}
\usepackage{amsmath} 
\usepackage{bm}
\usepackage{color}
\usepackage{subfig}
\usepackage{caption}
\usepackage{amssymb} 
\DeclareMathAlphabet{\mathcal}{OMS}{cmsy}{m}{n}
\newcommand{\be}{\begin{eqnarray}}
\newcommand{\bc}{\begin{center}}
\newcommand{\ec}{\end{center}}

\newcommand{\nn}{\nonumber}

\newcommand{\ex}[1]{\langle#1\rangle}
\newcommand{\bigex}[1]{\Big\langle#1\Big\rangle}
\newcommand{\ket}[1]{ |#1\rangle}
\newcommand{\bra}[1]{\langle #1|}

\usepackage{centernot}
\usepackage{cancel}

\begin{document}
\title{Entangled photon pair excitation and time-frequency-filtered multidimensional photon correlation spectroscopy as a probe for dissipative exciton kinetics}
\author{Arunangshu Debnath}
\email{arunangshu.debnath@desy.de}
\affiliation{Center for Free-Electron Laser Science CFEL, Deutsches Elektronen-Synchrotron DESY, Notkestrasse 85, 22607 Hamburg, Germany} 
\author{Shaul Mukamel}
\email{smukamel@uci.edu}
\affiliation{Department of Chemistry and Department of Physics and Astronomy, University of California, Irvine, California 92697, USA}


\begin{abstract}
In quantum aggregates, delocalized exciton states across energy manifolds interact with phonon modes, making state-resolved spectroscopic monitoring of dynamics challenging. We propose a scheme that combines photon-entanglement-enhanced narrowband excitation of two-exciton states with time-frequency-filtered two-photon coincidence counting, which allows high-resolution probing of dissipative two-exciton dynamics spread across multiple spectral and temporal windows. We demonstrate that entangled photon pairs can be used to prepare narrowband two-exciton population distributions, circumventing transport in the mediating one-exciton manifold, and the redistributed two-exciton population can be monitored using time-frequency-filtered two-photon coincidence counting.
Numerical simulations for a light-harvesting aggregate highlight the ability of this protocol to suppress or amplify specific pathways under a realistic scenario. Combining entangled photonic sources with multidimensional photon correlation techniques enable promising applications in spectroscopy and sensing.
\end{abstract}

\maketitle

\section{Introduction}\label{intro}
In interacting quantum molecular aggregates, optical probing of exciton kinetics and the extraction of state-resolved spectroscopic information are inherently difficult. The high density of exciton states and exciton-phonon interactions lead to dissipative dynamics that span multiple time and energy scales.
Molecular aggregates are found in naturally occurring photosynthetic systems.
Within the photosynthetic family, light-harvesting complexes are critical components that facilitate efficient photon energy conversion through exciton formation and transport. Investigating the structural and functional mechanisms of exciton dynamics provides essential insights for engineering biomimetic light-harvesting systems. \cite{duan2022quantum, abramavicius2009extracting, lorenzoni2025full}. 
Aggregation of individual chromophores results in delocalized electronic excitations of the Frenkel type, which participate in relaxation and dephasing, eventually leading to primary charge-separation processes.\\
Signatures of exciton-exciton interactions and correlations, which are typically hidden in the dynamics of multi-exciton states, have been investigated \cite{mueller2020molecular, maly2023separating, schroter2018using, debnath2026photonic}. The early attributes of such correlations are already evident in the energy distributions of two-exciton states, which deviate from the sum of two one-exciton energies. Previous studies have demonstrated that various modalities of two-photon absorption spectroscopy can monitor the associated resonances
\cite{spano1991biexciton, spano1991cooperative, gutierrez2021frenkel, you2015observation, kuhn1996two, yu2022double, debnath2013high}. \\
In recent years, several proposals focusing on atomic and simple molecular systems have highlighted the advantages of employing entangled photon pairs to probe two-exciton states \cite{schlawin2013suppression, javanainen1990linear, saleh1998entangled, pevrina1998multiphoton, dorfman2016nonlinear, ishizaki2020probing, debnath2020entangled, fujihashi2023probing, fujihashi2024pathway, raymer2013entangled, varnavski2023colors, malatesta2023optical, cho2018quantum, citeroni2025ultrafast}. These proposals were complemented by critical examinations of the role of photon entanglement in enhancing two-photon absorption \cite{raymer2022theory, landes2024limitations, landes2021quantifying, parzuchowski2021setting, raymer2021large, khan2024does, albarelli2023fundamental, das2025optimal} and in obtaining superior resolution in condensed-phase spectroscopies \cite{dorfman2019monitoring, debnath2020entangled}.
Alongside, previous studies have proposed novel two-photon fluorescence intensity correlation measurements to monitor the properties of generated two-exciton states \cite{lubin2022photon, pandya2024towards, dorfman2012nonlinear, dorfman2016time, sanchez2020photon}. In a series of theoretical works, time-frequency-filtered photon coincidence counting was proposed as a technique that may provide spectroscopic information similar to that of two-dimensional spectroscopy, but at different timescales \cite{dorfman2012nonlinear, dorfman2016time, dorfman2018multidimensional, zhang2018monitoring}.\\
The optical excitation of two-exciton states in molecular aggregates, such as light-harvesting complexes, typically occurs via a manifold of one-exciton states subjected to transport and dephasing. As a result, the population is redistributed among several states, such that only a few have a high propensity of transitions to the target two-exciton state. As the number of chromophoric sites increases, the number of states in the one and two-exciton manifolds increases as well. Consequently, the number of interfering optical transitions increases dramatically. Additionally, the generated two-exciton population also undergoes transport, quickly dispersing any initially created narrowband two-exciton distribution.
In this work, we demonstrate that two-photon excitation using entangled photon pairs can create a narrowband population distribution circumventing one-exciton transport. We also show that time-frequency-filtered two-photon counting can characterize the cascaded photon emission processes. The filtering parameters can be modulated to generate two-dimensional correlation plots, constituting a spectroscopic technique capable of monitoring the dissipative dynamics.\\
We focus on the light-harvesting complex LHCII, the most abundant antenna complex in plants and an important constituent of Photosystem II.
\begin{figure*}[ht]
\centering
  \includegraphics[width=.96\textwidth]{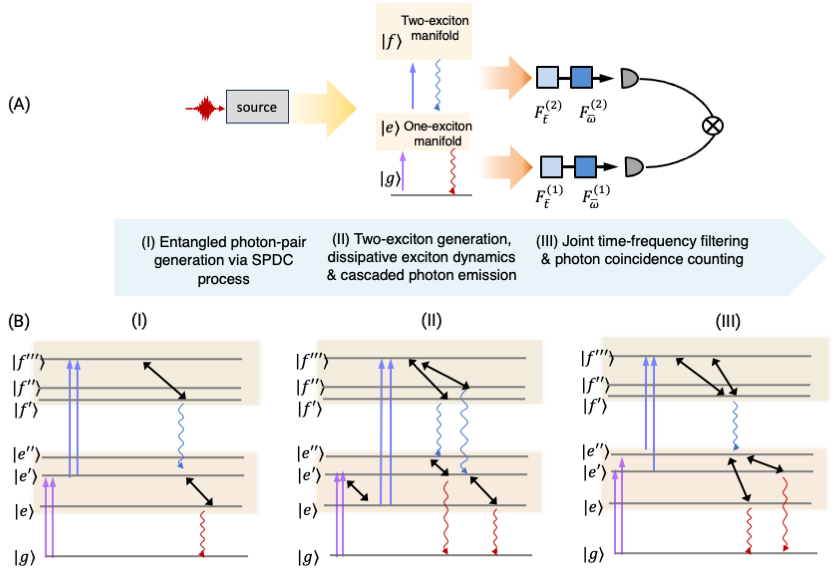}\hfill
\caption{(A) Illustration of the proposed protocol as a three-stage process: preparation of the two-exciton population, population redistribution, and time-frequency-filtered detection of cascaded two-photon emission. The action of the time filter, governed by the filtering function $F_{\bar{t}}^{(j)}$, precedes the frequency filter, governed by the filtering function $F_{\bar{\omega}}^{(j)}$, for both detection channels. Vertical straight arrows denote interactions with driving photon modes, whereas wiggly arrows denote interactions with the detector modes.  
(B) Representation of three exemplary dynamical pathways. Tilted double-ended arrows indicate intra-manifold exciton transport. (I) Two- and one-exciton transport preserves the narrowband nature of the initial population distribution, leading to narrowband emissions. (II) During the excitation stage, one-exciton transport alters the target two-exciton population distribution. Subsequent two-exciton transport causes population dispersal, resulting in broadband cascaded emission. (III) The broadband two-exciton population distribution funnels to a narrowband configuration, memory of which is erased by subsequent one-exciton transport; it leads to a narrowband emission followed by a broadband emission. Filtering functions assist in classification of the pathways and construction of a mechanistic narrative.}
  \label{fig:workflow}
\end{figure*}
It exists as a trimeric structure composed of structurally rigid monomeric aggregates; each monomer contains several chlorophylls whose lowest three energy levels are spectroscopically accessible. 
In Section~\ref{sec:ham}, we introduce the exciton-phonon Hamiltonian of the light-harvesting complex LHCII along with the exciton-photon Hamiltonian. In the subsequent sections, namely Sec.~\ref{sec:results} (\ref{subsec:entexcite}–\ref{subsec:pcc}), we detail the theoretical framework for the combined scheme and present numerical simulations. The manuscript concludes in Sec.~\ref{sec:conclude} with a summary of our findings and a brief commentary.
\section{Driven dissipative one and two-exciton kinetics}\label{sec:ham}
We simulate the one and two-exciton dynamics of the light-harvesting complex LHCII using the Frenkel exciton model within the Heitler-London approximation. The model is constructed by considering the lowest three electronic energy states for each chromophoric site. The Hamiltonian (with $\hbar=1$) is given by,
\begin{align}\label{eqn:ham}
    H_{}^{} 
    &=\sum_{mn} (E_m^{} \delta_{mn} + J_{mn}) B_m^\dag B_n^{}  \nn\\
    &+\sum_{mn} U_{m}^{(1)} B_m^\dag B_m^\dag B_m^{}B_m^{}+U_{mn}^{(2)} B_m^\dag B_n^\dag B_n^{}B_m^{}\nn\\
    &+\sum_{j}^{} \omega_j^{} (b^\dag_j b_j^{}+\frac{1}{2}) + \sum_{m,j}^{} g_{m,j}^{} (b^\dag_j+ b_j^{}) B_m^{\dag} B_m^{}
\end{align}    
where the exciton creation (annihilation) operators, $B_m^\dag$ ($B_m$), follow the commutation relation $[B_n, B_m^\dag] = \delta_{mn}(1 - \eta (B_m^\dag B_m)^2)$ (with $\eta = 3/2$) and the phonon creation (annihilation) operators, $b_j^\dag$ ($b_j$), follow the commutation relation $[b_i, b_j^\dag] = \delta_{ij}$.\\
In the above, $E_m$ and $J_{mn}$ describe the on-site excitation and inter-site hopping, respectively. The latter arises due to the Coulomb interaction and leads to the delocalization of the exciton states across the sites. The one-exciton Hamiltonian is obtained by considering the first term, i.e.,  $h_{mn}^{(1)}=(E_m^{} \delta_{mn} + J_{mn}) B_m^\dag B_n^{}$.
The two-exciton Hamiltonian is obtained by considering $h_{mnkl}^{(2)} = h_{mn}^{(1)} \delta_{kl} + \delta_{mn} h_{kl}^{(1)} +  U_{m}^{(1)} \delta_{mk} \delta_{nl} \delta_{mn}+ U_{mn}^{(2)} \delta_{mk} \delta_{nl}$
where $h_{mn}^{(1)}$ is given above. The matrix elements of the two-exciton Hamiltonian differ from the sum of two one-exciton Hamiltonians by the last two terms. These terms, which contribute to two-exciton nonlinearities, represent energetic shifts of overtone and combination exciton states, respectively. 
Diagonalization yields the corresponding one and two-exciton eigenstates, denoted $\ket{e_{j}^{}}$ and $\ket{f_{k}^{}}$, respectively, expressed as 
\begin{align}\label{eqn:stateproj}
    \ket{e_{j}^{}} &= \sum_m^{} T^{(1)}_{j^{}, m} B_m^\dag \ket{0} \nn\\
\ket{f_{k}^{}}&=\sum_{mn}^{} T^{(2)}_{k, mn} B_m^\dag B_n^\dag\ket{0}
\end{align}
Here, $T^{(1)}_{j,m}$ and $T^{(2)}_{k,mn}$ represent the matrix elements of the diagonalization matrices. They determine the contributions of individual sites to each eigenstate.\\
The parameters for the Frenkel exciton Hamiltonian were obtained from \cite{novoderezhkin2011intra, novoderezhkin2004energy, debnath2022entangled, debnath2026photonic, debnath2026photon}. 
For the LHCII complex, the number of chromophoric sites is $N_s = 14$, yielding $N_g = 1$ ground state, $N_e = 14$ one-exciton states, and $N_f = 105$ two-exciton states, consisting of $14$ overtone and $91$ combination states.
\begin{figure}[hb]
 \includegraphics[width=.48\textwidth]{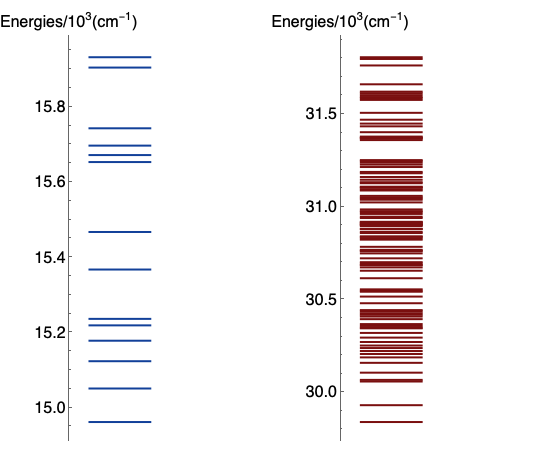}
\caption{One and two-exciton energies are displayed in the left and right panels, respectively. Nonuniform energy distributions are visible in both manifolds. Simulations in Section~\ref{subsec:entexcite} refer to the excitation of two-exciton states $f_{07}$ ($E_{f_{07}} =30192.9 \, \text{cm}^{-1}$) and $f_{83}$ ($E_{f_{83}} =31243.2 \, \text{cm}^{-1}$) via $e_{07}$ ($E_{e_{07}} =15307.1 \, \text{cm}^{-1}$) and $e_{09}$ ($E_{e_{09}} =15655.9 \, \text{cm}^{-1}$).}
  \label{fig:12exenergy}
\end{figure}
%
The one exciton manifold spans from $E_{e_{1}} = 14964.5 \,\text{cm}^{-1}$ to $E_{e_{14}} = 15934.7 \,\text{cm}^{-1}$, hosting $14$ states within a spectral width of $\approx 970 \,\text{cm}^{-1}$. The two-exciton manifold spans from $E_{f_{1}} = 29844.3 \,\text{cm}^{-1}$ to $E_{f_{105}} = 31811.8 \,\text{cm}^{-1}$, hosting $105$ states within a spectral width of $\approx 1968 \,\text{cm}^{-1}$. The corresponding energy distributions, plotted in Fig.~\ref{fig:12exenergy}, illustrate the nature of the density of states in each manifold.
The energetic ordering of exciton eigenstates is dependent upon two competing parameters: $J_{mn}$ that govern exciton delocalization and $U_{m}^{(1)}$, $U_{mn}^{(2)}$ that govern exciton nonlinearities \cite{debnath2026photonic}. \\
The dissipative exciton dynamics is described by the Liouville-space Green's function formalism. Using the exciton eigenbasis, we can express the Green's function as follows
\begin{align}\label{eqn:cohtransGF}
&\mathcal{G}_{a_1 a_2,a_3 a_4}^{}(t) = \delta_{a_1 a_2}\delta_{a_3 a_4} \theta(t) [\exp{(- K^{} t)}]_{a_1 a_1,a_3 a_3} \nonumber\\
&+ (1-\delta_{a_1 a_2}) \delta_{a_1 a_3}\delta_{a_2 a_4} \theta(t) \exp{(-i \omega_{a_1 a_3}^{} t -\gamma_{a_1 a_3}^{} t)}
\end{align}
The expression above is valid for both exciton transport and dephasing, which are captured by the first and the second terms, respectively. It is discussed further in Appendix~\ref{app:ham}.\\
Interaction with the photonic sources, within the rotating-wave approximation, is described by the following Hamiltonian:
\begin{align}\label{eqn:hamint}
   & H_{\mathrm{int}}^{}(t) = E V^\dag + E^\dag V =\sum_{j, pq \in \{ge,ef\}} \sqrt{\frac{2\pi \omega_j}{\Omega} }\nn\\&
     \Big( a_j^{} \exp{(-i \omega_j t)} 
    \exp{(i \omega_{pq}^{} t)}    B_{pq}
+\mathrm{h.c.} \Big)
\end{align}
Here, we used mode expansion for field operators $E$, where $a_j$, $\Omega$, and $\omega_j$ denote the photon annihilation operators, the mode volume, and the frequencies respectively. These mode operators follow Bosonic commutation relations. The exciton annihilation operators which include inter-manifold exciton transitions are denoted as $B_{pq}(t) =d_{pq}^{}\ket{p}\bra{q} (\text{with properties}\,\, \{p, q\} \in g,e,f$.\\
\begin{figure*}[ht]
\centering
  \includegraphics[width=.96\textwidth]{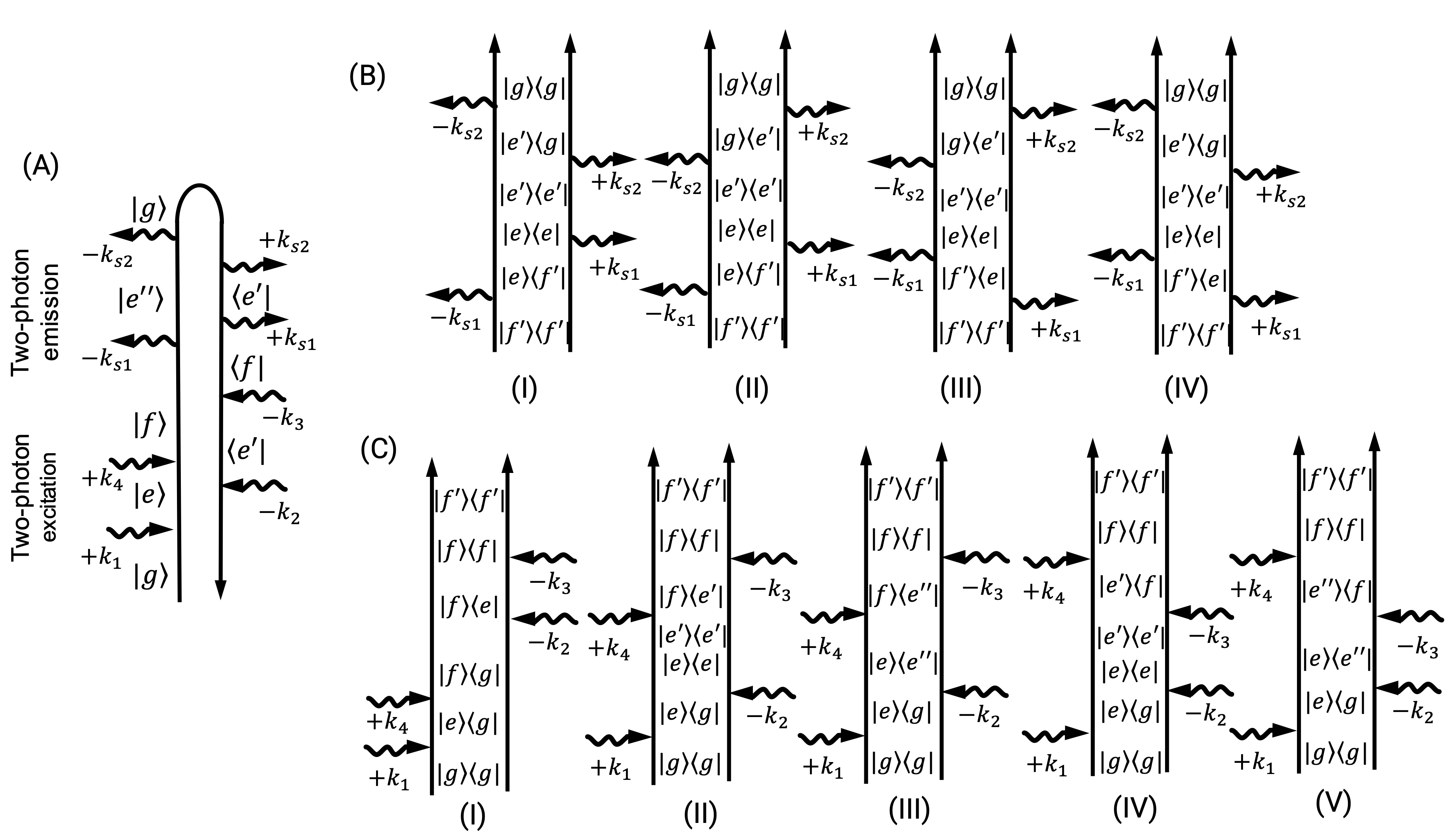}\hfill
\caption{(A) The Schwinger-Keldysh loop diagram describes eight exciton-photon interactions, including four exciton-photon interactions each with the driving (denoted $k_j$) and detection modes (denoted $k_{s_j}$). For dissipative dynamics, equivalent Liouville space ladder diagrams, (B) and (C), are suitable. Section \ref{sec:results} presents the corresponding equations. Panel (C) describes the preparation stage. Pathway (I) depends on inter-manifold two-exciton coherence, $\ket{f}\bra{g}$; pathways (III) and (V) depend on intra-manifold one-exciton coherence, $\ket{e}\bra{e''}$; and the pathways in (II) and (IV) depend on one-exciton transport, $\ket{e}\bra{e} \rightarrow \ket{e'}\bra{e'}$. Panel (B) describes the detection stage. All pathways originate from the population distribution following two-exciton transport, $\ket{f}\bra{f} \rightarrow \ket{f'}\bra{f'}$. A combinatorial pairing of these two sets of pathways accounts for all possible contributions to the signal. Their relative contributions are controlled by the four-point correlation functions of the entangled photon pairs and the time-frequency filtering properties. }
\label{fig:diagram}
\end{figure*} 
We consider entangled photon pairs generated via spontaneous parametric down-conversion (SPDC) in the weak-down-conversion limit. In a typical scenario, a birefringent crystal is pumped with a classical laser, leading to the generation of correlated photon pairs with frequencies $\omega_1$ and $\omega_2$. The SPDC process involves two photon-matter interactions with the incoming classical field modes and four photon-matter interactions with the outgoing signal-idler fields. The correlation properties of the emitted photon field are determined by the physical properties of the SPDC crystal. \cite{saleh1998entangled, schlawin2018entangled, javanainen1990linear, schlawin2013suppression, dorfman2014multidimensional, debnath2020entangled, debnath2022entangled}.\\
The generated entangled two-photon states is described as follows:
\begin{align}
\rho_{\text{photon}} &=
\int\frac{d\omega_{a}}{2\pi}
\int\frac{d\omega_{b}}{2\pi} 
\int\frac{d\omega_{a}'}{2\pi}
\int\frac{d\omega_{b}'}{2\pi} \nn\\&
F(\omega_{a},\omega_{b}) F_{}^{*}(\omega_{a}',\omega_{b}') 
a_{\omega_a}^{\dag}a_{\omega_b}^{\dag} \ket{0}\bra{0} a_{\omega_a'}^{}a_{\omega_b'}^{}
\end{align}
The four-point correlation function of the entangled photon pairs, which is relevant for the creation of two-exciton states, is evaluated using the expression above, yielding 
\begin{align}\label{eqn:entfieldfactor}
    &\ex{E^{\dag}_{}(\omega_a')E^{\dag}_{}(\omega_b')
    E^{}_{}(\omega_b)E^{}_{}(\omega_a) }_{\rho_{\text{photon}}}\nn\\&
    = F_{}^{*}(\omega_a',\omega_b') F_{}^{}(\omega_a,\omega_b)
\end{align}
where
$F_{}^{}(\omega_{a}^{},\omega_{b}^{})  = \alpha A_p^{}(\omega_{a}^{}+\omega_{b}^{}) 
\Big\{ \mathrm{sinc}[\phi(\omega_a,\omega_b)] \exp{i\phi(\omega_a, \omega_b)} +a \leftrightarrow b \Big\}$, the $\phi(\omega_a,\omega_b) =(\omega_a^{}-\omega_1^{})\tilde{T}_1^{}/2 
+(\omega_b^{}-\omega_2^{})\tilde{T}_2^{}/2$; $\alpha$ denotes the parameter governing the down-conversion efficiency.
The entanglement time parameter $\tilde{T}_{\mathrm{ent}}=\tilde{T}_2-\tilde{T}_1$ establishes an upper bound for the intervals between two photon-exciton interactions and depends on the time delays acquired by the signal and idler photons during propagation within the nonlinear crystal. It characterizes the non-classical correlation in the photonic source.  
The classical transform-limited pump pulse driving the SPDC crystal is defined as follows: $A_p^{}(\omega_{}^{}) = E_0 \sqrt{\pi/\Gamma_{p,0}}\exp{[-(\omega-\omega_{p})^2/ 4 \Gamma_{p,0}]}$, where $\Gamma_{p,0}=1/T_{0}^{2}=2 \ln 2/\tau_{0}^{2}$ and $T_{0}$  the temporal half-width at $1/e$. $\tau_{0}$ and $\omega_{p}$ is the temporal width and the central frequency of the pump, respectively.
The Joint Spectral Amplitude $F(\omega_a, \omega_b)$ is governed by two characteristic scales: diagonal width $\Delta\omega_+$ which is related to the duration of the pump pulse $\Gamma_{p,0} \propto 1/\Delta\omega_+$, and the counter-diagonal width $\Delta\omega_-$ which is related to the entanglement time $\tilde{T}_{\mathrm{ent}} \propto 1/\Delta\omega_-$. The absolute values are displayed in Fig.~\ref{fig:entcorr81}.
A two-dimensional Fourier transform yields the joint temporal amplitude function: $\tilde{\Psi}(\tau_a, \tau_b)$, which is parametrized by center-of-mass and relative time coordinates: $t_{+} = (\tau_a + \tau_b)/2$ and $t_{-} = \tau_a - \tau_b$ satisfying: $\Delta t_{+} \propto 1/\Delta\omega_{+}$ and $\Delta t_{-} \propto 1/\Delta\omega_{-}$, respectively. The quantities $\Delta\omega_+$ and $\Delta\omega_-$ are governed independently at the level of SPDC process by the pump laser and nonlinear crystal dispersion/length.
\section{Results}\label{sec:results}
We depict the sequence of events, entangled photon excitation, dissipative evolution, and two-photon emission, occurring in three stages, in Fig.~\ref{fig:workflow} (A) (I) to (III). 
Excitation by entangled photon pairs creates a tailored two-exciton population distribution which undergoes transport for a finite time. This results in a redistribution of the populations. The redistributed population serves as the initial configuration for the cascaded two-photon emission processes. The emission process involves two successive one-photon emission events interleaved by one-exciton transport in the one-exciton manifold.\\
In Fig.~\ref{fig:workflow} (B), we demonstrate three plausible pathways via which the dynamics may proceed. Due to transport, the two-exciton population distribution may lose its narrowband character or even become narrower. The dynamical information is encoded into the photon emissions that correspond to the inter-manifold transitions between the upper two manifolds ($\ket{f} \rightarrow \ket{e} $). The one-exciton transport that follows may also create additional dispersion of the one-exciton population. The corresponding dynamical information is encoded into the inter-manifold transition between lower two manifolds ($\ket{e} \rightarrow \ket{g} $).
These two sets of photon emissions can be distinguished by controlling the properties of the time and frequency filters, denoted $F_{\bar{t}}^{(j)}$ and $F_{\bar{\omega}}^{(j)}$ in the Fig.~\ref{fig:workflow}. \\
In the subsequent sections, we denote Liouville-space operators: $A_\alpha$, (where $\alpha = L, R, +, -$), defined by their action on a Hilbert-space operator $X$ as $A_L X \equiv A X$ and $A_R X \equiv X A$. The corresponding matrix elements are obtained as $(A_\alpha X)_{ij} \equiv \sum_{kl} (A_\alpha)_{ij, kl} X_{kl}$. We also define: $A_+ \equiv \frac{1}{2}(A_L + A_R)$ and $A_- \equiv A_L - A_R$, where the matrix elements are given by: $(A_{-})_{ij,kl} = A_{ik}
\delta_{jl} - A_{\ell j} \delta_{ik}$ and
$(A_{+})_{ij,kl} = \frac{1}{2}[A_{ik} \delta_{jl}
+ A_{l j} \delta_{ik}]$.\\
The time-frequency filtered two-photon coincidence signal is expressed as
\begin{align}\label{eqn:sigfull}
& S(\overline{t}_{2},\overline{\omega}_{2};
\overline{t}_{1},\overline{\omega}_{1}) 
=\int_{-\infty}^{\infty}dt_{2}' d\tau_{2} 
\int_{-\infty}^{\infty}dt_{1}' d\tau_{1} \nn\\&
\sum_{s_{2},s_{2'},s_{1},s_{1'}} \text{Tr}_{s_{2},s_{2'},s_{1},s_{1'}}
\big[D^{(s_{2},s_{2'})}(\overline{t}_{2},\overline{\omega}_{2};t_{2}',\tau_{2}) \nn\\&
D^{(s_{1},s_{1'})}(\overline{t}_{1},\overline{\omega}_{1};t_{1}',\tau_{1}) 
B^{(s_{2},s_{2'},s_{1},s_{1'})}(t_{2}',\tau_{2};t_{1}',\tau_{1})\big]_{}
\end{align}
where we defined the following operators. The operator describing time-frequency filtered photo-detection, $D^{(\nu,\nu')}(\overline{t}_{\nu},\overline{\omega}_{\nu};t_{\nu}',\tau_{\nu})$ is given by
\begin{align}
& D^{(s_{\nu},s_{\nu'})}(\overline{t}_{\nu},\overline{\omega}_{\nu};t_{\nu}',\tau_{\nu}) \nn\\
& =D(\overline{t}_{\nu},\overline{\omega}_{\nu};t_{\nu}',\tau_{\nu})
E_{\nu'L}^{}(t_{\nu}') E_{\nu R}^{\dagger}(t_{\nu}'+\tau_{\nu})
\end{align}
and operator describing two-photon emission, 
$B^{(s_{2},s_{2'},s_{1},s_{1'})}(t_{2}',\tau_{2};t_{1}',\tau_{1}) $, is given by
    \begin{align}\label{eq:sigpccbare}
    &   B^{(s_{2},s_{2'},s_{1},s_{1'})}(t_{2}',\tau_{2};t_{1}',\tau_{1}) \nn\\
  &= T_{L}^{(s_{2'},s_{1'})}(t_{2}',t_{1}')
 T_{R}^{(s_{2},s_{1})\dagger}(t_{2}'+\tau_{2},t_{1}'+\tau_{1}) 
    \end{align}
where
\begin{align}
  T_{R}^{(s_{2},s_{1})\dagger}(t_{2}'+\tau_{2},t_{1}'+\tau_{1})  &= i^2 
    \int_{-\infty}^{t_2'+\tau_2} dt_4 \int_{-\infty}^{t_1'+\tau_1} dt_3 \nn\\&
    E_{s_{2} R}(t_3)E_{s_{1} R}(t_4) V_{R}^{\dag}(t_4) V_{R}^{\dag}(t_3) \nn\\
   T_{L}^{(s_{2'},s_{1'})}(t_{2}',t_{1}') &= i^2 
    \int_{-\infty}^{t_2'} dt_2 \int_{-\infty}^{t_1'} dt_1 \nn\\&
    E_{s_{2'} L}^\dag (t_1)E_{s_{1'} L}^\dag(t_2) V_{L}^{}(t_1) V_{L}^{}(t_2)
\end{align}
The trace is taken over the final two-exciton population following transport, $\rho_{f'f'}^{}(t') = \mathcal{G}_{f'f',ff}(t'-t) \rho_{ff}^{} (t)$; it serves as the initial state for the photon emission events following the entangled photon pair excitation.
The two-exciton population $ \rho_{ff}^{} (t)$ is obtained by using the projection operator $\ket{f}\bra{f}$ acting on the two-exciton density operator, i.e., $\rho_{ff}^{} (t) = \text{Tr}[O(t)\ket{f}\bra{f}]_{}$, where 
\begin{align}\label{eqn:sigrhoff}
  & O(t) = \tilde{T}_{R}^{\dag} (\tau_2,\tau_3)\tilde{T}_{L}^{}(\tau_1,\tau_4) = i^2 
    \int_{-\infty}^{\tau_2} d\tau_4 \int_{-\infty}^{\tau_1} d\tau_3   \nn\\&
    \int_{-\infty}^{\tau_3} d\tau_2 \int_{-\infty}^{\tau_4} d\tau_1 
   E_{L}^\dag (\tau_4)E_{L}^\dag(\tau_1) E_{R}(t_3)E_{R}(\tau_2) \nn\\&
  \qquad \qquad  V_{R}^{\dag}(\tau_3)  V_{R}^{\dag}(\tau_2)  V_{L}^{}(\tau_4) V_{L}^{}(\tau_1)
\end{align}
and the partial trace is taken over the phonon modes. The operator $\tilde{T}_{\nu}^{\dag} (\tau_i,\tau_j)$ define entangled two-photon excitation process in the perturbative limit.
The discussion above assumed that the excitation and emission stages do not overlap temporally, which resulted in the factorization of Eq.~\ref{eqn:sigfull} into three separate contributions: preparation, evolution and detection. In the following, we present the key microscopic expressions and numerical simulations analyzing each of them.
%
\begin{figure*}[ht!]
 \includegraphics[width=.96\textwidth]{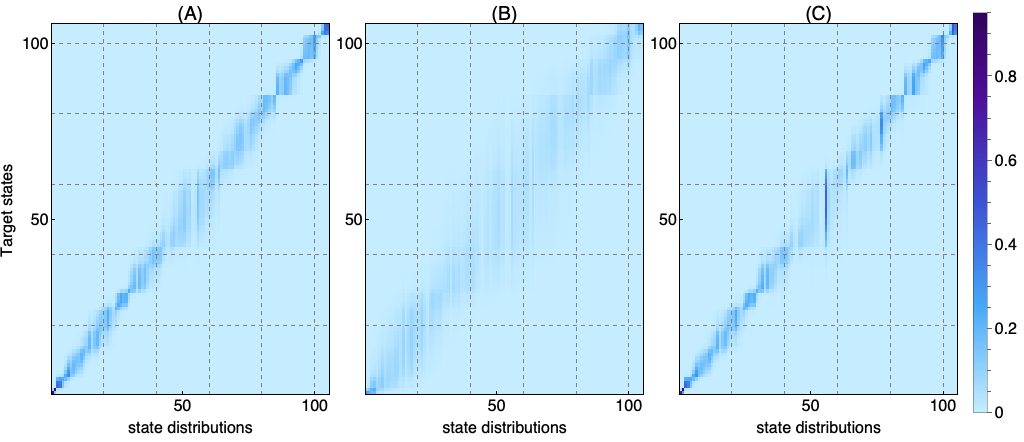}
\caption{Two-exciton population distribution map, constructed by sequentially targeting each of the $105$ states in the manifold using degenerate entangled photon pairs ($\omega_i = \omega_p/2 = E_{\text{target}}^{(2)}/2$). The left panel serves as a reference simulation where an entanglement time of $T_{\mathrm{ent}}=10 \, \mathrm{fs}$ and an SPDC pump width of $\tau_0 = 150 \, \mathrm{fs}$ is used. In the middle and right panels, the temporal width of the SPDC pump and the entanglement time are varied, respectively using $\tau_0 = 50 \, \mathrm{fs}$) and $T_{\mathrm{ent}}=60 \, \mathrm{fs}$, while keeping all other parameters unchanged. While taking a horizontal cut, the dispersal away from diagonal signifies a broadband excitation. For a detailed discussion, see Section~\ref{subsec:entexcite}}
  \label{fig:popall}
\end{figure*}
\subsection{Entangled photon-pair excitation}\label{subsec:entexcite}
The quantity of interest at the preparation stage, the two-exciton population distribution, is contained in $\rho_{ff}^{} (t) $, defined in Eq.~\ref{eqn:sigrhoff}. It is evaluated (following the steps outlined in Appendix~\ref{app:signalderiv}, and using the Green's functions defined in Eq.~\ref{eqn:cohtransGF}) as,
\begin{widetext}
\begin{align}\label{eqn:pop1compact}
\rho_{ff}^{} (t)
& \propto 2 \text{Re}\Big\{ \int \frac{d\omega_4^{}}{2 \pi}\int \frac{d\omega_3^{}}{2 \pi}\int \frac{d\omega_2^{}}{2 \pi}\int \frac{d\omega_1^{}}{2 \pi} 
e^{+ i\omega_3^{} \tau_3^{} +i\omega_4^{} \tau_4^{} - i\omega_2^{} \tau_2^{}- i \omega_1^{} \tau_1^{} }
\ex{E^{\dagger}(\omega_4)E^{\dagger}(\omega_3)E(\omega_2)E(\omega_1)}\nn\\
&
\Big\{d_{fe'}^{}d_{e'g}^{}d_{fe}^{*}d_{eg}^{*}  
\times I(-\omega_4^{}-\omega_3^{}+\omega_2^{}+\omega_1^{}; - z_{f f}^{})
 I(-\omega_3^{}+\omega_2^{}+\omega_1^{} ;- z_{f e'}^{})
I(+ \omega_2^{}+\omega_1^{};  - z_{fg}^{})
I(\omega_1^{}; - z_{eg}^{} )  \nn\\
&+ d_{fe'}^{}d_{fe}^{*} d_{e'g}^{}d_{eg}^{*}\times
 I(-\omega_4^{}+\omega_2^{}- \omega_3^{}+\omega_1^{}; - z_{f f}^{})
I(\omega_2^{}- \omega_3^{}+\omega_1^{} ;-z_{fe^{''}}^{})\nn\\
&  \Big( \delta_{e e'} \sum_p \chi^{\text{R}}_{e'' p} D_{pp}^{-1}I(- \omega_3^{}+\omega_1^{};- z_p^{}) \chi^{\text{L}}_{p e'} + (1 - \delta_{e e'}) \delta_{e e''} I(- \omega_3^{}+\omega_1^{}; - z_{e e'}^{}) \Big)
 I(\omega_1^{}; - z_{eg}^{})\nn\\
& +d_{fe'}^{}d_{fe}^{*} d_{e'g}^{}d_{eg}^{*}\times
 I(+\omega_2^{}-\omega_4^{}- \omega_3^{}+\omega_1^{}; - z_{f f})
 I(-\omega_4^{}- \omega_3^{}+\omega_1^{}; - z_{fe''})\nn\\
& \Big( \delta_{e e'} \sum_p \chi^{\text{R}}_{e'' p} D_{pp}^{-1}
I(- \omega_3^{}+\omega_1^{};- z_p^{}) \chi^{\text{L}}_{p e'} 
+ (1 - \delta_{e e'}) \delta_{e e''} 
I(- \omega_3^{}+\omega_1^{}; - z_{e e'}^{}) \Big)
I(\omega_1^{}; - z_{eg}^{}) \Big\}\Big\} 
\end{align}
\end{widetext}
Each of the five terms described above represents a distinct exciton pathway contributing to the signal. The pathways include contributions from both exciton coherences and populations. The first term represents the two-exciton coherence-mediated pathway (diagram (I)), which is susceptible to inter-manifold dephasing. The second and fourth terms describe pathways susceptible to one-exciton population transport (diagrams (II) and (IV)). The third and fifth terms describe pathways susceptible to intra-manifold dephasing (diagrams (III) and (V)). The Liouville-space Feynman diagrams in Fig.~\ref{fig:diagram} (I)–(V), offer further insights into the exciton processes captured by them.\\
The expression above is particularly useful for exploring the role of entangled photon pairs; their effect is included via the four-point correlation functions defined in Eq.~\ref{eqn:entfieldfactor}. The following sections investigate the roles of both degenerate and nondegenerate entangled photon pairs.
\subsubsection{Simulation: excitation of the full suite of two-exciton states using degenerate entangled photon pairs}\label{subsubsec:entexcite1}
In this section, we monitor the final two-exciton population distribution following excitation by spectrally degenerate entangled photon pairs, i.e., $\omega_1 = \omega_2 = \omega_p/2 = E_{\mathrm{target}}^{(2)}/2$. For an SPDC source pumped by a finite-bandwidth laser, the generated photon pairs are spread over a range of frequencies; they must be filtered to select the degenerate components.
We carry out a parametric scan, taking each of the $N_{f}=105$ two-exciton states as a target, and present the final normalized two-exciton population distribution in Fig.~\ref{fig:popall}. The results presented in the left column correspond to an entanglement time of $T_{\mathrm{ent}}=10 \, \mathrm{fs}$ and an SPDC pump width of $\tau_0 = 150 \, \mathrm{fs}$, which acts as a reference case. The middle and right columns present simulations with a variation in the temporal width of the SPDC pump ($\tau_0 = 50 \, \mathrm{fs}$) and the entanglement time parameter ($T_{\mathrm{ent}}=60 \, \mathrm{fs}$), respectively, while keeping all other parameters unchanged.\\
In all three cases, when the target states are located near the middle of the two-exciton manifold, the population distributions are broad around the target, indicating a lower degree of selectivity. When the target states are located in the lower-energy sector of the two-exciton manifold, a narrow population distribution indicates a higher degree of selectivity.
Altering the temporal width of the SPDC pump pulse, as shown in the middle column, resulted in diminished selectivity, likely due to the activation of multiple excitation pathways. In comparison, altering the entanglement time in the right column preserved the selectivity for the majority of states. Across all plots, we notice a trend: some two-exciton states remain preferentially populated across all parameter regimes. This feature has likely emerged from the role of one-exciton transport parameters of mediating states and requires further investigation. 
\begin{figure*}[ht!]
 \includegraphics[width=.96\textwidth]{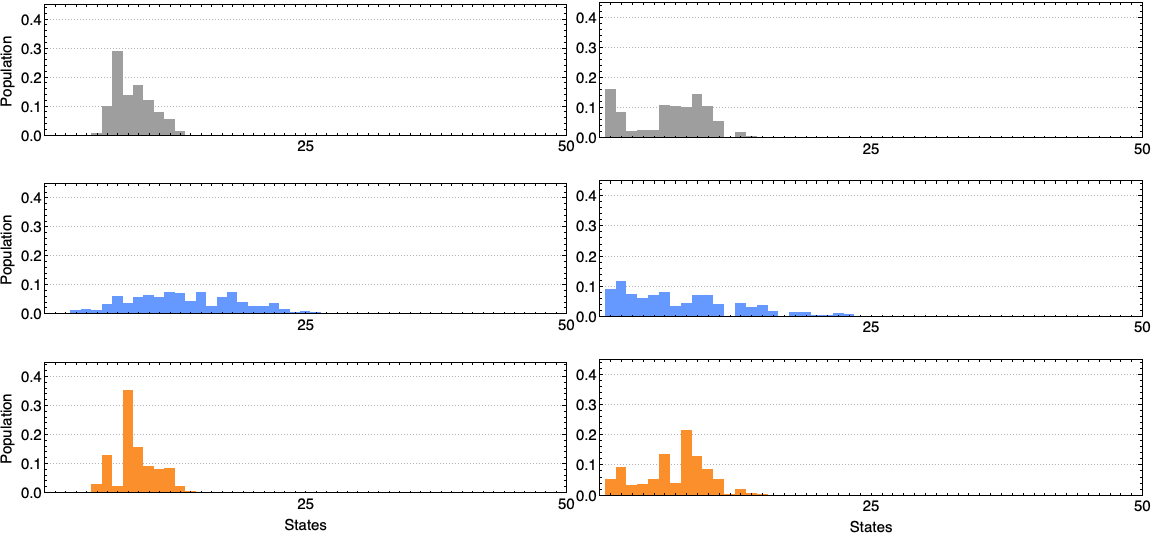}
  \caption{The population distribution following the targeted excitation of $f_{07}^{}$ via the intermediate states $e_{07}^{}$ and $e_{09}^{}$ is shown for two durations of transport: $\tilde{\tau} =0\, \text{fs}$ (left column) and $\tilde{\tau} =50 \, \text{fs}$ (right column). The upper panel displays the distribution for an entanglement time of $\tilde{T}_{\text{ent}} =10 \, \text{fs}$ and an SPDC pump width of $\tau_0= 150 \, \text{fs}$. The middle and lower panels show the distributions when the width is reduced to $\tau_0= 50 \, \text{fs}$ and the entanglement time is increased to $\tilde{T}_{\text{ent}} =30 \, \text{fs}$, respectively. For a detailed discussion, refer to Section~\ref{subsec:entexcite}.} 
  \label{fig:pop07evolve}
\end{figure*}
%
\subsubsection{Simulations: differential preparation of two-exciton populations via preselected mediating states and parametric dependence on photonic properties}\label{subsubsec:entexcite2}
Among the full suite of target states previously monitored, we focus on two specific ones: $f_{07}^{}$ and $f_{83}^{}$, residing in the lower ($E_{f_{07}} =30192.9 \, \text{cm}^{-1}$) and the higher ($E_{f_{83}} =31243.2 \, \text{cm}^{-1}$) energy regions of the two-exciton manifold, respectively. We investigate the possibility of exciting them via a fixed pair of one-exciton states: $e_{07}$ and $e_{09}$, which are prone to delocalization and ultrafast transport. They are selected by setting $\omega_1= E_{e_{07}} =15307.1 \, \text{cm}^{-1}$ and $\omega_2=E_{e_{09}} =15655.9 \, \text{cm}^{-1}$.\\
The results are displayed in the left columns of Fig.~\ref{fig:pop07evolve} and Fig.~\ref{fig:pop83evolve}, respectively. The entanglement time and temporal width of the SPDC pump are set to $T_{\mathrm{ent}}=10 \, \mathrm{fs}$ and $\tau_0 = 150 \, \mathrm{fs}$ for the upper panel, which is similar to the reference case in Fig.~\ref{fig:popall}. In the middle panel, the temporal width is changed to $\tau_0 = 50 \, \mathrm{fs}$, whereas in the bottom panel, the entanglement time is changed to $T_{\mathrm{ent}}=40 \, \mathrm{fs}$.\\
The results demonstrate that the desired narrowband excitation is indeed possible, albeit with low final occupancy in some cases. The increase in the temporal width results in a significant loss of selectivity as the number of excitation pathways with comparable spectral weights grows. The variation of the entanglement time did not result in any major changes, except for the tailoring of the relative populations. The latter, however, will have consequences as the population redistributes over time, as will be shown in the next section. We emphasize that both the target states are excited via the same mediating one-exciton states. Therefore, the differential nature of the final population distribution is dependent on the transition dipoles that connect chosen one-exciton states to the two-exciton target states. The final population distribution demonstrates that the short entanglement time reaches saturation, less so for the $f_{07}$ than $f_{83}$. The latter is a higher-energy state and corresponding transitions impose stringent frequency requirements.\\ 
Unlike the systems with few intermediate states, the number of competing pathways in the LHCII is large, which limits the ability of entangled photon pairs to achieve perfect target excitation. Long $\tilde{T}_{\mathrm{ent}}$, in our case, has two effects: it allows longer transport and opens up competing transitions that degrade the success probability for the target state. Such competing transitions may also be present, albeit in smaller numbers, in the case short $\tilde{T}_{\mathrm{ent}}$.
\subsection{Transport in the two-exciton manifold}\label{subsec:poptransport}
The created two-exciton population distribution evolves over time, governed by the kinetic equation:
\begin{align}\label{eqn:poptransportkinetic}
   \frac{d}{dt}\rho_{ff}(t)=-\sum_{f'}K_{ff,f'f'}\rho_{f'f'}(t)  
\end{align}
where the transport matrix, $K_{ff,f'f'}$, follows the detailed balance condition $K_{f'f',ff }/K_{ff,f'f'} =\exp(-\omega_{f'f}/(k_{B}T)) $, and $\sum_{f}K_{ff,f'f'}=0$. The transport matrix dictates the nature of the population redistribution. The initial two-exciton population distribution, obtained via Eq.~\ref{eqn:pop1compact}, is propagated using the expression $\rho_{ff}(t_2) = \mathcal{G}_{ff,f'f'}(t_2-t_1)\rho_{f'f'}(t_1)$, where $\mathcal{G}_{ff,f'f'}(\tilde{\tau})$ is the transport Green's function introduced in Eq.~\ref{eqn:cohtransGF} and $\tilde{\tau}=t_2-t_1$, the duration of transport; it is calculated from the kinetic equation above, as described in Appendix~\ref{app:ham}.\\
This raises several questions.: How does the initial distribution redistribute over time in the two cases investigated? To what extent do the two different initial distributions differ following this redistribution? Finally, how long does it take before the memory of the initial distribution, which depends on the properties of the entangled photon source, is lost? We investigate these questions via simulations below.
\\
\subsubsection{Simulations: differential evolution of two-exciton populations and parametric dependence on photonic properties}\label{subsubsec:poptransdifferent}
\begin{figure*}[ht!]
 \includegraphics[width=.96\textwidth]{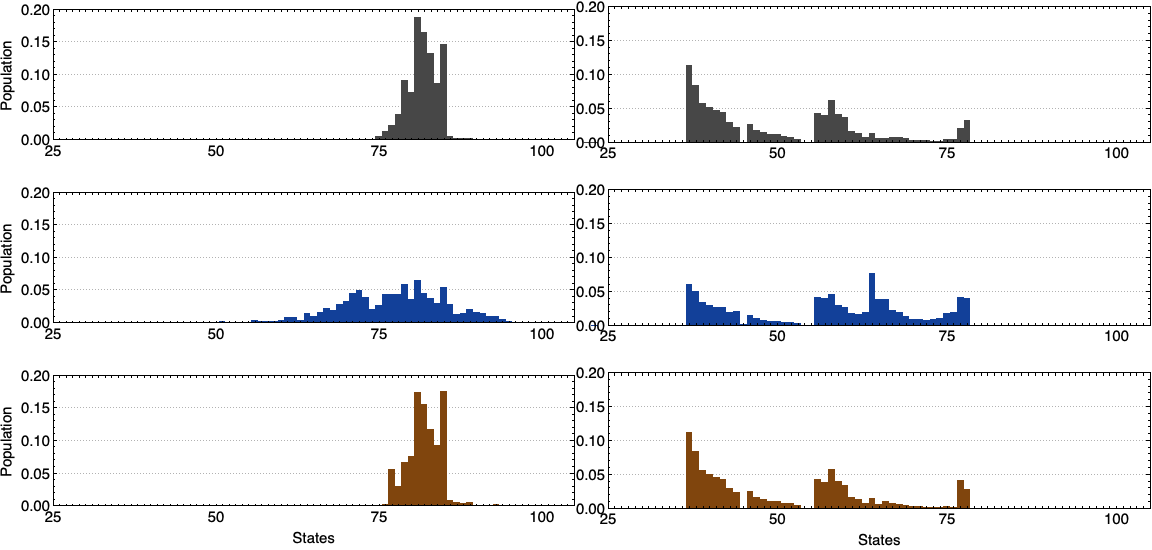}
 \caption{Population distribution for target state $f_{83}$, which is dominantly excited via intermediate states $e_{07}$ and $e_{09}$, is shown for two durations of transport, $\tilde{\tau} =0 \,\text{fs} $ (left column), and $\tilde{\tau} =250 \,\text{fs} $ (right column). In the upper panels, the entanglement time is $\tilde{T}_{\text{ent}} = 10\,\text{fs} $ and the SPDC pump width is $\tau_0 = 150\,\text{fs} $. In the middle panels, the width is changed to $\tau_0 = 50\,\text{fs} $. In the lower panels, the entanglement time is increased to $\tilde{T}_{\text{ent}} = 30\,\text{fs} $. Section~\ref{subsec:entexcite} provides a detailed discussion of these excitation dynamics.}
  \label{fig:pop83evolve}
\end{figure*}

We monitor the temporal evolution of populations following targeted excitation of $f_{07}$ (right column, Fig.~\ref{fig:pop07evolve}) and $f_{83}$ (right column, Fig.~\ref{fig:pop83evolve}), at durations of transport $\tilde{\tau} = 50\, \text{fs} $ and $\tilde{\tau} = 250\, \text{fs} $, respectively.
Each of the three panels, upper, middle, and bottom, presents simulations using the same parameters used in the reference one in Section~\ref{subsubsec:entexcite2}.
While the distributions in Fig.~\ref{fig:pop07evolve} retain the memory of the prepared state at short times, this memory is erased in Fig.~\ref{fig:pop83evolve}. The rapid redistribution and localization toward a few selected states in the $f_{83}$ manifold is a key feature that requires further investigation. The influence of varying the photonic parameters is less pronounced in Fig.~\ref{fig:pop83evolve}, indicating that the distribution loses the primary advantages afforded by entangled photon pair excitations.
\subsubsection{Simulations: differential redistribution of two-exciton populations}\label{subsubsec:poptransredis}
\begin{figure*}[ht!]
 \includegraphics[width=.96\textwidth]{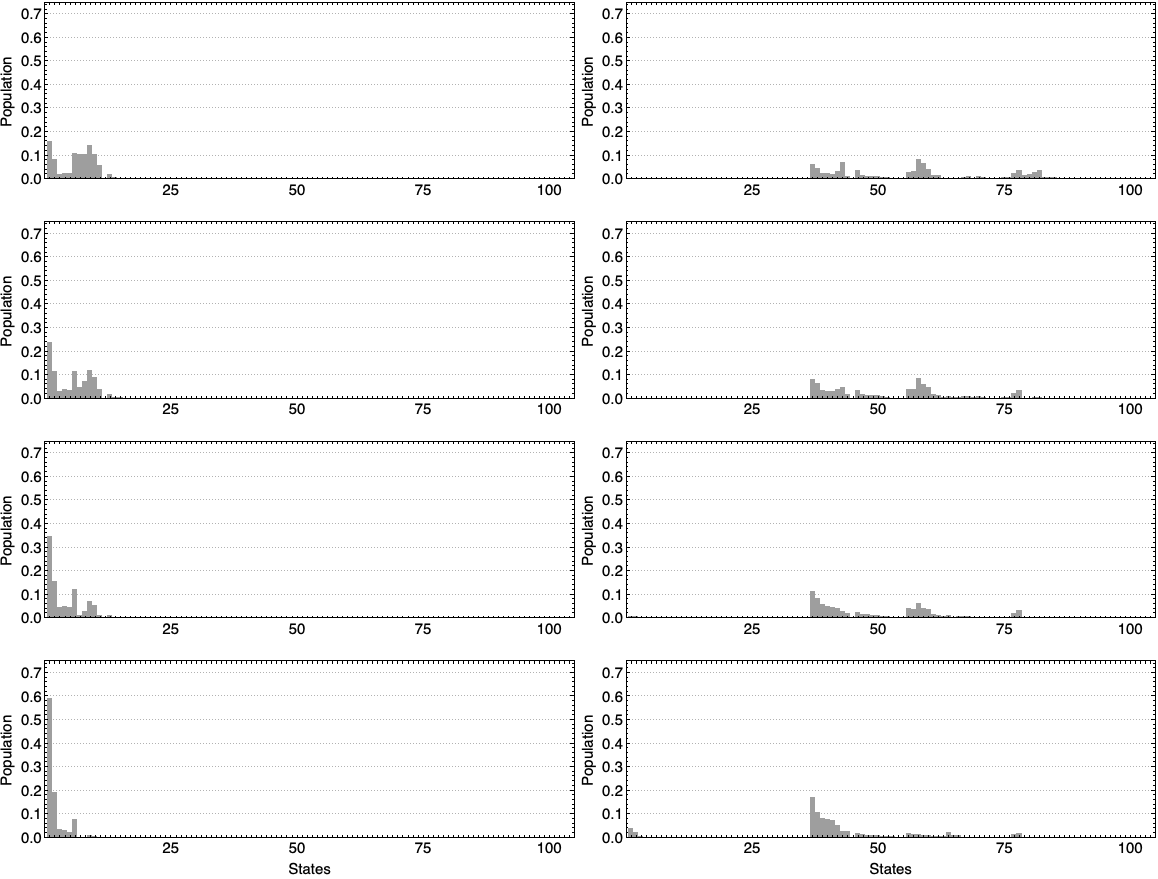}
\caption{Population redistribution following the excitation of target states $f_{07}$ (left column) and $f_{83}$ (right column) plotted for $\tilde{\tau}= 50,\, 100,\, 250$ and $1000 \,\text{fs}$. The results demonstrate two key features: rapid population migration that follows energy gradients at longer times for all and cascaded migration for the higher lying target state. Section~\ref{subsec:poptransport} provides a detailed discussion. } 
  \label{fig:pop07propagate}
\end{figure*}

In Fig.~\ref{fig:pop07propagate}, we track the population redistribution following the targeted excitation of $f_{07}$ (left column) and $f_{83}$ (right column). We monitor population snapshots taken at four durations of transport: $\tilde{\tau} = 50$, $100$, $250$, and $1000 \, \text{fs}$.
The results demonstrate two distinct features: for $f_{07}$, the population monotonically migrates to the lower-energy sector of the manifold, whereas for $f_{83}$, it migrates via a slower intra-manifold cascading mechanism. The slow transport is attributed to a lack of overlap between $f_{83}$ and the states in the middle rung of the two-exciton manifold. This observation generalizes the findings shown in Figs.~\ref{fig:pop07evolve} and \ref{fig:pop83evolve}.\\
\subsection{Photon correlation spectroscopy}\label{subsec:pcc}
The quantity of interest in photon correlation spectroscopy is the time–frequency filtered two-photon counting signal. The population redistribution is usually tracked by monitoring two-photon emissions that accompanies cascaded inter-manifold transitions. We propose that the joint time-frequency filtering of the two-photon counting signal \cite{dorfman2014multidimensional, dorfman2019monitoring, mukamel2011multidimensional} can track the population in a state- and time-resolved manner. 
The rationale behind the filtering scheme can be developed as follows. Population transport in the two-exciton manifold acts to redistribute the initially created population. The timescale of this phenomenon is determined by the duration of transport. The population redistribution has two principal consequences: it gives rise to a new set of inter-manifold pathways and reduces the spectral intensity of some transitions. Each photon emission pathway is associated with a specific time and frequency window, which can be monitored by tuning the properties of the first set of filters.
Following photon emission, the dynamics continue in the one-exciton manifold, where transport may scramble the population distribution. After a finite duration, another inter-manifold transition leads to the ground state. Similar to the above, these transitions can also be monitored via the second set of filters. 
Three pathways in Fig.~\ref{fig:workflow} (B) show the dynamical complexities that accompany each photon emission. In systems with exciton manifolds, multiple photon emission pathways are possible.\\
The signal is expressed as: 
\begin{align}\label{eqn:pcc1}
& S(\bar{t}_2,\bar{\omega}_2; \bar{t}_1,\bar{\omega}_1) =\int dt_1' \int d\tau_1 D^{(1)}(t_1,\omega_1;t_1',\tau_1)\nn\\&
\int dt_2' \int d\tau_2 D^{(2)}(t_2,\omega_2;t_2',\tau_2)
\bigex{N_{}(t_1',\tau_1) N_{}(t_2',\tau_2)} 
\end{align}
Here, $N_{}(t',\tau) =\sum_{s_{\nu},s_{\nu'}} E_{s_{\nu}^{} R}^{\dagger}(t'+\tau)E_{s_{\nu'}^{}L}(t')$ is a photon number operator without the filter. The field operators for the photonic modes of the detector can be expanded as: $E_s(t)=\sqrt{2\pi\omega_s/\Omega_s} a_s e^{-i\omega_s t}$, where $\Omega_s$ is the mode quantization volume. The field operators follow the Bosonic commutation relation
$[a_s, a_{s'}^\dagger] = \delta_{ss'}$.
Here, we used a simplified form of the functions describing filtered photodetection, $D^{(j)}(\bar{t}_j, \bar{\omega}_j; t_j', \tau_j)$, which are further governed by the time and frequency filtering functions, $F_{\bar{t}_j}^{}$ and $F_{\bar{\omega}_j}^{}$, respectively. It takes the form
\begin{align}\label{eqn:Ddef}
& D^{(j)}(\bar{t}_j,\bar{\omega}_j,t',\tau) \nn\\&=
\int\frac{d\omega''}{2\pi}e^{-i\omega''\tau}| F_{\bar{\omega}_j}(\omega'',\omega)|^2 F_{\bar{t}_j}^{*}(t'+\tau,t) F_{\bar{t}_j}(t',t).
\end{align}
For time and frequency filters, we choose the following functional form: $ F_{\bar{t}_j}(t')=\theta(t'-\bar{t}_j) 
\exp({ - \sigma_{t_j} (t'-\bar{t}_j)}) $ and $
   F_{\bar{\omega}_j}(\omega' )= i(\omega'-\bar{\omega}_j+i \sigma_{\omega_j})^{-1}$.
The function is evaluated as follows
\begin{align}\label{eqn:Dlor}
& D^{(j)}(\bar{t}_j,\bar{\omega}_j,t',\tau)  =\frac{1}{2\sigma_{\omega_j}}\theta(t'-\bar{t}_j)\theta(t'+\tau-\bar{t}_j) \nn\\&
[\theta(\tau)e^{-\sigma_{\bar{\omega}_j}\tau}
+\theta(-\tau)e^{\sigma_{\bar{\omega}_j}\tau}] 
e^{-(i\omega+\sigma_{\bar{t}_j}^{})\tau-2\sigma_{\bar{t}_j}^{}(t'-\bar{t}_j)}
\end{align}
It depends on four parameters: the centering times and frequencies of the time and frequency filters, denoted $\bar{t}_j$ and $\bar{\omega}_j$, and their respective widths, $\sigma_{\bar{t}_j}$ and $\sigma_{\bar{\omega}_j}$. The expressions for the filtering functions contain Heaviside functions that select the detection times.
The frequency filters resolve inter-manifold transitions that originate from exciton transport occurring within similar time windows. The time filters help separate the transport processes that may involve similar spectral gaps. Together, these parameters monitor inter-manifold transitions jointly in time and frequency space.\\
Using the expression above and following the steps described in Appendix~\ref{app:gatingderiv}, we express the signal as a sum of four pathways (which is reduced to two following symmetry arguments) as follows
\begin{widetext}
   \begin{align}\label{eqn:pccpath12}
&S_{}^{}(\bar{t}_2,\bar{\omega}_2; \bar{t}_1,\bar{\omega}_1)
=\sum_{p=1}^{4} S_{}^{(p)}(\bar{t}_2,\bar{\omega}_2; \bar{t}_1,\bar{\omega}_1) 
=2 \text{Re}\sum_{p=1}^{2}S_{}^{(p)}(\bar{t}_2,\bar{\omega}_2; \bar{t}_1,\bar{\omega}_1)\nn\\&
=\sum_{g,e,e',f,f'} |\mathcal{D}(\bar{\omega}_1)\mathcal{D}(\bar{\omega}_2)|^2|d_{eg}|^2|d_{ef}|^2 \times\int dt_1' \int d\tau_1 D_>^{(1)}(\bar{t}_1,\bar{\omega}_1; t_1',\tau_1)
\times\int dt_2' \int d\tau_2 D_>^{(2)}(\bar{t}_2,\bar{\omega}_2; t_2',\tau_2) \nn\\&
\langle \bra{I}\mathcal{G}_{ge',ge'}(\tau_1)\mathcal{G}_{e'e',ee}(t_1'-t_2'-\tau_2)\mathcal{G}_{ef',ef'}(\tau_2)\mathcal{G}_{f'f',ff}(t_2') \ket{\rho_{ff}}\rangle \nn\\&
+\sum_{g,e,e',f,f'} |\mathcal{D}(\bar{\omega}_1)\mathcal{D}(\bar{\omega}_2)|^2|d_{eg}|^2|d_{ef}|^2 \times\int dt_1' \int d\tau_1 D_>^{(1)}(\bar{t}_1,\bar{\omega}_1; t_1',\tau_1)\times\int dt_2' \int d\tau_2 D_<^{(2)}(\bar{t}_2,\bar{\omega}_2; t_2',\tau_2) \nn\\&
\langle \bra{I} \mathcal{G}_{ge',ge'}(\tau_1)\mathcal{G}_{e'e',ee}(t_1'-t_2')\mathcal{G}_{f'e,f'e}(-\tau_2)\mathcal{G}_{f'f',ff}(t_2'+\tau_2)\ket{\rho_{ff}}\rangle 
\end{align} 
\end{widetext}
Here, subscripts($>, <$) in the expression of $D^{(j)}(\bar{t}_j, \bar{\omega}_j; t_j', \tau_j)$ indicate the sign of the delay $\tau$ between two photon emissions. The density of photonic states of the detector, $\mathcal{D}(\bar{\omega}_1)\mathcal{D}(\bar{\omega}_2)$, is assumed to be peaked at centering frequencies of the filter.\\ 
We illustrate the photon emission pathways using Liouville space diagrams in Fig.~\ref{fig:diagram} ((I)–(IV)); each diagram corresponds to pathways that involve transport in both manifolds, duration of which is controlled by the centering times of the time filters.
\begin{figure*}[ht!]
 \includegraphics[width=.96\textwidth]{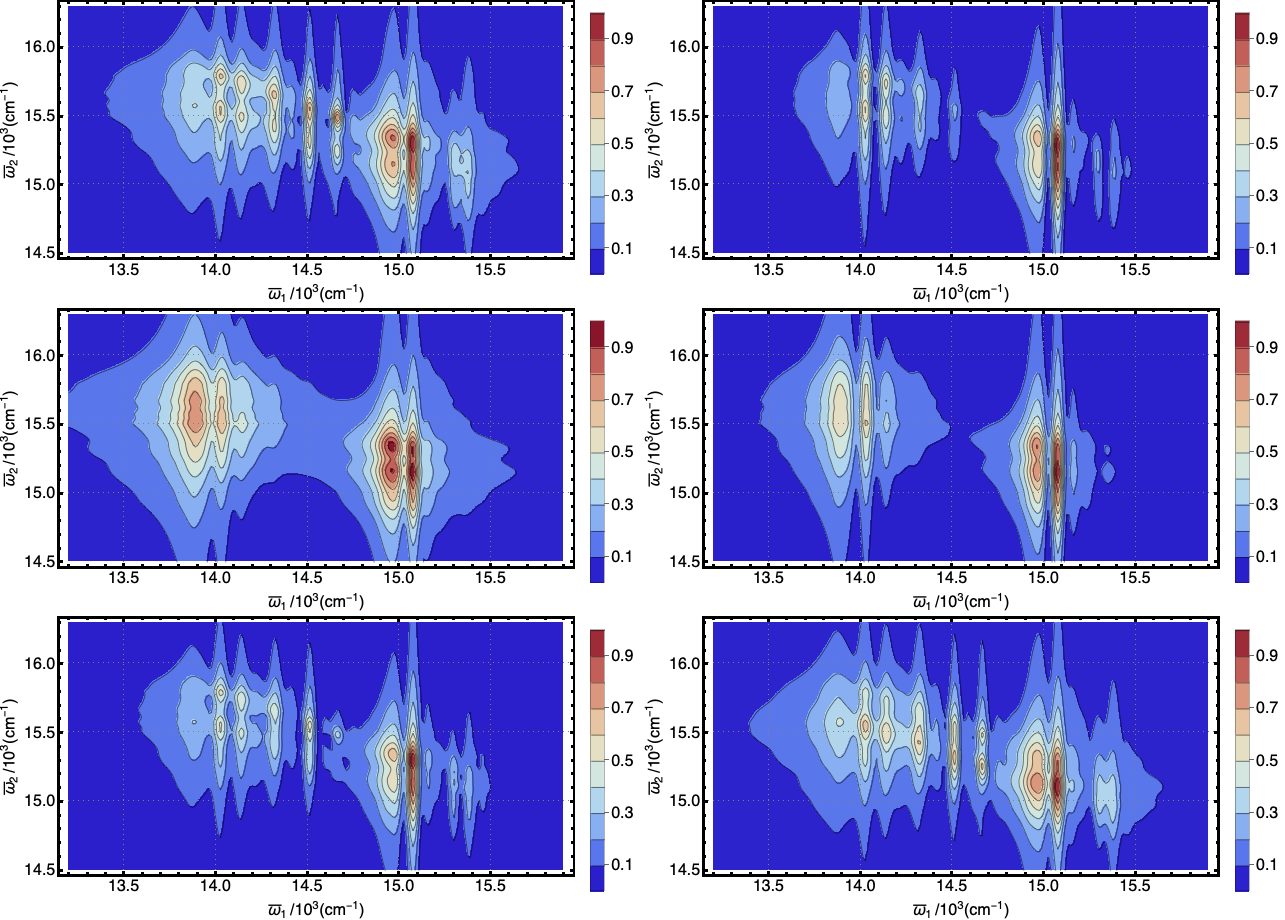}
  \caption{
The time-frequency-filtered two-photon counting signal is shown. The initial state is defined by the population distribution targeted at $f_{07}$, followed by two-exciton transport. 
  The reference simulation in the upper-left uses the following parameters: frequency filter widths $\sigma_{\bar{\omega}_1} = \sigma_{\bar{\omega}_2}= 10\,\text{cm}^{-1}$, time filter widths $\sigma_{\bar{t}_1} = \sigma_{\bar{t}_2}= 4.8681\,\text{cm}^{-1}$, and waiting times $t_{\text{w},1} = 0\,\text{fs}$, and $t_{\text{w},2} = 100\,\text{fs}$. In the upper-right panel, the frequency filter width is increased to $\sigma_{\bar{\omega}_1} = \sigma_{\bar{\omega}_2}= 20\,\text{cm}^{-1}$. 
  In the middle-left panel, the centering time of the time filter is adjusted to $t_{\text{w},2} = 1000$ fs. In the middle-right panel, the frequency filter widths are changed to $\sigma_{\bar{\omega}_1} = \sigma_{\bar{\omega}_2}= 20\,\text{cm}^{-1}$ while maintaining the same centering as the middle-left.
  In the bottom-left panel, the time filter width is changed to $\sigma_{\bar{t}_2}= 0.5409\,\text{cm}^{-1}$.
  Finally, in the bottom-right panel, the centering time is adjusted to $t_{w,1} = 50 \text{fs}$.
Section~\ref{sec:resultspcc} provides a detailed discussion of these results.}. 
  \label{fig:pcc}
\end{figure*}
\subsubsection{Simulation: two-photon coincidence counting signal with variation in filtering parameter and different waiting times}\label{sec:resultspcc}
In Fig.~\ref{fig:pcc}, we present six two-dimensional correlation plots by taking the centering times of the frequency filters, $\bar{\omega}_j$ as the scanning parameters. Peaks along the $\bar{\omega}_1$ axis correspond to transitions from the two-exciton manifold to the one-exciton manifold ($\ket{f} \rightarrow \ket{e}$), while peaks along the $\bar{\omega}_2$ axis correspond to transitions from the one-exciton manifold to the ground state ($\ket{e} \rightarrow \ket{g}$). The former monitors spectral gaps $\omega_{fe}$, whereas the latter monitors gaps $\omega_{eg}$, provided that the centering time of time filter $\bar{t}_2$ is ahead of $\bar{t}_1$.\\ 
The redistributed population following the excitation of target state $f_{07}$ serves as the initial state for the photon emission. As highlighted in Section~\ref{sec:ham} and discussed in Appendix~\ref{app:excitonenergy}, the density of spectral gaps, $\omega_{fe}$, is high. For $f_{07}$, the two-exciton manifold hosts $\sim 10$ states within a spectral width of $350\,\text{cm}^{-1}$. Given that the one-exciton manifold hosts $14$ states within a spectral width of approximately $970\,\text{cm}^{-1}$, we expect a maximum of $\sim 140$  transitions. Modulating the filtering parameters within realistic parameter bounds helps classify many of them.\\
The reference simulation (upper-left panel) uses  $\sigma_{\bar{\omega}_1} = \sigma_{\bar{\omega}_2}= 10\,\text{cm}^{-1}$, for widths of the frequency-filter, and $\sigma_{\bar{t}_1} = \sigma_{\bar{t}_2}= 4.8681\,\text{cm}^{-1}$ for widths of the time-filter.
We define variables $t_{\text{w},j}$, which are associated with the $\tau_j$ integrals in Eq.~\ref{eqn:pccpath12}; they determine the duration of effective durations of exciton transport in respective manifolds before the time filter is applied. We use $t_{\text{w},1} = 0\,\text{fs}$, and $t_{\text{w},2} = 100\,\text{fs}$. 
In the plot, we observe eight zones along the $\bar{\omega}_1 $ axis, where each of them can be further resolved along the $\bar{\omega}_2 $ axis. For easier interpretation, the peaks are classified into four distinct clusters along $\bar{\omega}_1/10^3\,\text{cm}^{-1}$: the first cluster contains two resonances ($\sim 14.0 - 14.2$), the second three ($\sim 14.3 - 14.7$), the third two ($\sim 14.9 - 15.1$), and the fourth one ($\sim 15.3 - 15.5$). These peaks, originating from phonon-broadened spectral gaps, would have appeared in a congested form in the absence of the time filter. \\
The upper-right panel repeats the reference simulation by increasing the widths of the frequency filter to $\sigma_{\bar{\omega}_1} = \sigma_{\bar{\omega}_2}= 20\,\text{cm}^{-1}$, while keeping all other parameters fixed. Along the $\bar{\omega}_1$ axis, several peaks in the second cluster are absent; the most noticeable absences occur in the middle and low energy sectors. Additionally, some peaks along $\bar{\omega}_2$ have vanished, and the fourth cluster exhibits diminished intensity. Since the transitions have originated from a limited number of two-exciton states preferentially transitioning to few one-exciton states, it indicates that broadband frequency filters increase the number of destructive interferences.\\
The middle-left panel investigates the effect of longer effective durations of one-exciton transport by setting $t_{\text{w},2} = 1000\,\text{fs}$, while keeping all other parameters unchanged from the reference simulation. 
The population redistribution in the one-exciton manifold depends on two factors: the dipole-connectivity with the two-exciton manifold and the transport rates. We observe that along the $\bar{\omega}_1$ axis, the peaks have further coalesced, giving rise to two dominant clusters. Additionally, the fourth cluster has vanished, and the resolution along the $\bar{\omega}_2$ axis is lost. Since at longer waiting times, the connectivity dictated by transition dipoles focuses the spectral weights to a few selected states, we see drastic reduction in resolution along both axis.\\ 
The middle-right panel repeats the reference simulation but with $\sigma_{\bar{\omega}_1} = \sigma_{\bar{\omega}_2}= 20\,\text{cm}^{-1}$ and $t_{\text{w},2} = 1000\,\text{fs}$, examining the combined role of broadband frequency filters and longer effective durations of one-exciton transport. 
We observe that the widening of the spectral gate did not permit a larger number of transitions, as corroborated by saturation of peaks along $\bar{\omega}_2$. This stems from two factors: first, the initial two-exciton distributions are restricted to a limited set of states with no further pathways to explore; second, the transitions at longer times involve dynamics that converge to this same set of states.\\
The bottom-left panel examines the role of the width of the time filter by setting $\sigma_{\bar{t}_2} = 0.5409  \, \text{cm}^{-1} $ while
keeping all other parameters identical to the reference simulation. We observe that the second cluster (in the upper-left) is now significantly diminished, and several resonances along the $\bar{\omega}_2$ axis are absent. Centering the filter at earlier times effectively excludes the resonances generated by long-term transport. Therefore, a comparison between this simulation and the reference identifies the transitions that occur at longer times.\\
The bottom-right panel explores the role of finite effective durations of two-exciton transport on the reference simulation by setting $t_{\text{w},1} = 50  \, \text{fs}^{} $. We observe that along $\bar{\omega}_2$, the double peaks have coalesced into single ones, while no marked change is observed along the $\bar{\omega}_1$ axis. The two-exciton transport, at longer times, funnels the population into a few low-lying two-exciton states (Fig.~\ref{fig:pop07evolve} and Fig.~\ref{fig:pop07propagate}), which reduces the number of possible $\ket{f} \rightarrow \ket{e}$ transitions. For developing a phenomenological description, the state indices from the previous set of figures need to be converted to energy values.
\section{Conclusions}\label{sec:conclude}
In this work, we demonstrated that combining entangled photon-pair excitation and filtered two-photon counting provides a route to classify dissipative two-exciton kinetics in interacting molecular aggregates. The light-harvesting complex II (LHCII) served as a testbed. The roles of exciton–exciton interactions and exciton–phonon interactions, hidden in the energy distribution of two-exciton states and the associated energy gaps, are reflected in the kinetics of population redistribution. Employing time and frequency filters allows spectroscopic probing with higher resolution than bare two-photon counting.\\
Results in Section~\ref{subsec:entexcite} demonstrate that high-fidelity excitation of selected two-exciton states is possible despite one-exciton transport in the mediating manifold. We examined the robustness of the narrowband population distribution against variations in the entangled photon parameters and the temporal width of the SPDC pump. We demonstrated the differential nature of the excitation process by exciting two distinct target states via the same one-exciton states; the analysis offers insight into pathway selectivity via tuning non-degenerate entangled photon frequencies. From these simulations, a clear strategy emerges: preferentially accessing the one-exciton resonances that are less susceptible to transport is the key to achieving robust, state-selective excitation of two-exciton states.\\
The expression in Eq.~\ref{eqn:pop1compact} remains applicable for a wide range of photonic sources, for example, squeezed states two-photon, thermal and stochastic photonic sources. Derived assuming a perturbative regime of exciton–photon coupling, the expression consists of five terms, each of which was evaluated from a convolution of four-point exciton and photon correlation functions. We demonstrated that the photonic correlation functions can be used to modulate spectral weights for each exciton pathway. Our future communication will address how multiobjective optimization algorithms can be used to actively tailor these photon correlation functions. \\
The evolution of the created two-exciton population distribution, which serves as the initial configuration for the photon emission pathways, is monitored in Section~\ref{subsec:poptransport}.
The inter-manifold two-exciton transport effectively scrambles the identity of the states created by two-photon excitation. Interestingly, the one-exciton transport exerts a similar scrambling effect on two occasions: during the excitation and again during the emission process. It impacts the interpretability of inter-manifold radiative transitions, which are already complicated by exciton transport that spans multiple time and energy scales. \\
In Section~\ref{subsec:pcc}, we deployed two-pairs of time and frequency filters to classify inter-manifold transitions and generate two-dimensional correlation maps. It is established that variation of the filtering parameters can be used to obtain insights into the distribution of spectral gaps, intra-manifold transport timescales, and inter-manifold dephasing. The filtering effectively enhances the resolution of the bare two-photon counting signal. The two-dimensional correlation maps may eventually allow one to construct a kinetic description of state-resolved dissipative exciton dynamics. The expressions in Eq.~\ref{eqn:pccpath12} remain valid in general. We derived the expression assuming that the excitation and emission processes are temporally separated which neglects spontaneous Raman contributions. Although the filtering parameters are independent variables governing the resolution, their variations often have concomitant effects on the resolution of the peaks. While presenting these variations, the near-term realizability of the parameters was considered.\\
Measuring one-exciton populations, both as residuals during the excitation and emission processes provides distinct dynamical information. In the former case, the population serves as an estimate for the success of the entangled photon pair excitation. In the latter case, the population allows the radiative emission pathways to be correctly assigned. We note that the coherence mediated two-photon excitation pathways are the ones dominantly affected by the photon entanglement.
The two-dimensional maps presented in Fig.~\ref{fig:pcc} provides a protocol for monitoring the one-exciton population; the marginal intensity distribution along the vertical axis is the key quantity that, for a given set of temporal widths of the time gates, is proportional to the one-exciton population. To examine the residual one-exciton population during the entangled photon pair excitation, one must focus on short-time fluorescence or resort to interferometric detection techniques. This will be addressed in future communication.\\
The SPDC process considered in this work is driven by a pulsed Gaussian laser field. 
The generation of entangled photon pairs is a stochastic process; the signal and idler photons can be generated anywhere along the length of the crystal. Once generated at a given position, they propagate to the exit face of the crystal at their respective group velocities, which determine their mean arrival times at the crystal output face and are related to the entanglement time. The pump pulse duration sets the time window during which the entangled photons are generated. 
The Fourier transform of the joint spectral amplitude function yields the joint temporal amplitude, which provides further insights into the interaction times with the excitonic system under investigation \cite{grice1997spectral, mikhailova2008biphoton, jeronimo2009control, maclean2018direct}. 
Also, for a pulsed SPDC process with nondegenerate signal and idler frequencies, maximal entanglement is no longer achievable. In typical experiments, selecting the frequencies of interest involves an additional filtering stage prior to interaction with the matter. This will be discussed in a future communication. 
Due to the Fourier uncertainty relation, broadband (ultrashort) pulses can suppress transport within the intermediate manifold but lose spectral selectivity among adjacent final states $\ket{f_k}$. For narrowband pulses, higher selectivity comes at the expense of longer one-exciton transport, which can cause almost all population in the excited state $\ket{e_j}$ to relax into lower $\ket{e_{j'}}$ prior to the second photon interaction, suppressing favorable $\ket{e_{j'}} \rightarrow \ket{f_k}$ pathways. We assume that entangled photon-pair generation is separated from the onset of photon-counting (\ref{fig:workflow}). \\
The Markovian simulations presented in this article can be extended to accommodate both non-Markovian dynamics and driving-induced dissipation \cite{debnath2013dynamics, debnath2012chirped}. In particular, non-Markovian algorithms, such as the hierarchical equations of motion (HEOM), auxiliary density operators, or tensor-train techniques involve significantly higher computational costs. For a given $N$-site exciton aggregate, there are $N_e =N$ one-exciton states and $N_f = N + N(N-1)/2$ two-exciton states, which lead to an exciton density operator with dimensions $(N_e+N_f)^2 \times (N_e+N_f)^2$. The model requires $1$ overdamped Brownian oscillator mode for low-energy phonons, and $48$ structured Brownian oscillator modes for high-energy phonons, and $10$ Matsubara modes to approximate the thermal distribution functions; all of these variables increase the size of the auxiliary density matrices that are co-propagated along with the exciton density operator. Implementing any of these non-Markovian methods remains a computationally challenging yet promising direction.\\
In general, entangled photon pairs offer two primary advantages for nonlinear multiphoton spectroscopies: the ability to modulate competing excitation (and de-excitation) pathways to selectively suppress or amplify specific exciton resonances, and more favorable scaling of the signal with the intensity of the incoming field.
The entanglement time, controlled by the length of the birefringent crystal, determines the temporal window of successive photon-exciton interactions. In contrast, the bandwidth of the classical spontaneous parametric down-conversion (SPDC) pump determines the range of accessible resonances. The fact that these variables can be controlled independently affords additional flexibility. 
With increasing size of the molecular aggregate, the number of exciton states increases, and the one-exciton kinetics may span a broader range of time and frequency values. In such cases, a robust excitation of two-exciton states would require a broader tunability of photonic sources \cite{christ2011probing, mauerer2009colors, debnath2023theory}. To characterize a wide range of two-photon emission pathways, similar constraints apply to the filtering parameters as well.\\
The correlation function of driving optical fields in our case, for coherent laser sources, factorize into the product of four amplitudes, i.e., $\langle E^{\dagger}(\omega_4) E^{\dagger}(\omega_3) E(\omega_2) E(\omega_1) \rangle = A^{*}_4(\omega_4) A^{*}_3(\omega_3) A_2(\omega_2) A_1(\omega_1)$. The signal scales with the square of the intensity of the incoming field, in contrast to a signal using entangled photon pairs, which scales linearly. Due to the difficulty of separating contributions arising from photonic entanglement and inherently low signal-to-noise ratio, identifying parameter regimes in which photonic entanglement provides practical benefits remains a challenge \cite{pollmann2025limitations, khan2024does, kizmann2023quantum}. Technological advances in developing intense entangled photon sources would help establish the proposed protocol as a tool for investigating non-classical effects in light-harvesting complexes. \\
The time-frequency-filtered $n$-photon counting signal can be used to analyze multiphoton statistics in a $2n$-dimensional parameter space. If combined with a multiphoton entanglement-enhanced state preparation technique, it provides a robust spectroscopic scheme for characterizing dissipative molecular excitonic dynamics in large molecular systems, even when they span multiple time and frequency scales.
The proposed approache can also be extended to include programmable photonic sources and interferometric detection schemes \cite{krenn2021conceptual, dorfman2021hong, asban2021distinguishability, zhang2026theory, adamou2025quantum}. 
\begin{acknowledgments}
A.D. acknowledges the support from DESY (Hamburg, Germany), a member of the Helmholtz Association HGF. S.M. gratefully acknowledges the support of the National Science Foundation (NSF) grant Grant No. CHE-2246379.
\end{acknowledgments}
\appendix
\section{The Hamiltonian}\label{app:ham}
In the Hamiltonian, the Eq.~\ref{eqn:ham}, the vibrational terms are specified in the third line. The normal modes associated with collective intermolecular vibrations constitute the phonon modes. The latter, in the continuum limit, $J_{\mathrm{ph}}(\omega)= \sum_{j}  |g_{j}^{}|^2 [\delta(\omega-\omega_{j})-\delta(\omega+\omega_{j})]$ are described using an overdamped oscillator and $48$ multimode Brownian oscillators. The corresponding spectral function is given by
\begin{align}\label{eqn:phononspectralden}
J_{\mathrm{ph}}(\omega)&=2\lambda_0 \frac{\omega\gamma_0}{\omega^2+\gamma_0^2}+ \frac{2\lambda_j \omega_j^2 \omega \gamma_j}{(\omega_j^2-\omega^2)^2+\omega_{}^2\gamma_j^2}  
\end{align}
Here, $\lambda_0$ and $\gamma_0$ represent the damping strength and relaxation parameter for the overdamped mode, respectively. The Brownian oscillators are characterized by the Huang-Rhys parameter, mode frequency, and correlation time, denoted by $\lambda_j$, $\omega_j$, and $\gamma_j$, respectively. Relevant parameter values are detailed in the Ref.~\cite{debnath2022entangled, debnath2020entangled, novoderezhkin2011intra, novoderezhkin2004energy}.\\
The phonon interaction gives rise to exciton transport (first term) and dephasing (second term) Eq.~\ref{eqn:cohtransGF}. 
The exciton transport Green's function can be expressed as 
\begin{align}
\mathcal{G}_{a_1 a_1,a_3 a_3}^{}(t) &=[\exp{(- K^{} t)}]_{a_1 a_1,a_3 a_3} \nn\\
&= \sum_p \chi_{a_1 p}^R D_{pp}^{-1}e^{-\lambda_p t} \chi_{pa_3}^L
\end{align}
where $\lambda_p$ is the p-th eigenvalue of the transport matrix, $\chi^L (\chi^R)$ are the left (right) eigenvectors that satisfy $ K \chi^R =\lambda_p \chi^R$ and $ \chi^L K  =\lambda_p \chi^L$ of the same with additional relation $D_{}=\chi^L\chi^R$.\\
The intra-manifold exciton coherence Green's function is given by
\begin{align}\label{eqn:gfdephasing}
    \mathcal{G}_{a_1a_2}(t) &= \exp{[-i\omega_{a_1a_2} t-\gamma_{a_1a_2}^{} t]}
\end{align}
where we defined the pure dephasing parameter
\begin{align}\label{eqn:paramdephasing}
\tilde{\gamma}_{a_1a_2}^{}=\gamma_{a_1a_2}^{}-\frac{1}{2}(K_{a_1a_1,a_1a_1}+K_{a_2a_2,a_2a_2}).
\end{align}
and the dephasing parameters $\gamma_{a_1a_2}^{} = - K_{a_1a_2,a_1a_2}$.
%
\begin{figure}[ht]
 \includegraphics[width=.48\textwidth]{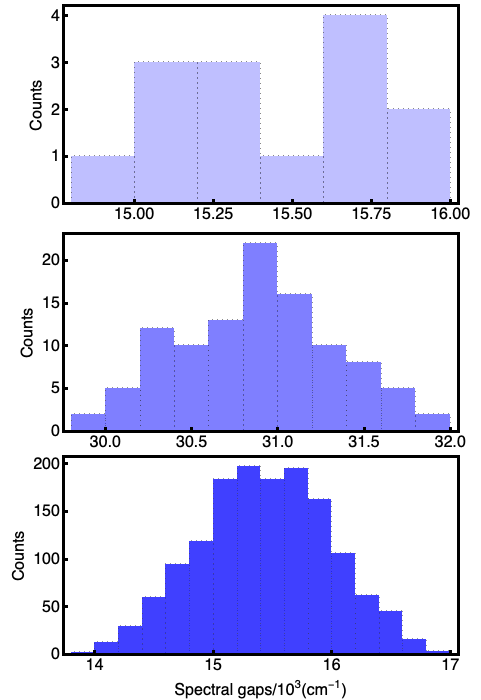}
\caption{Histograms illustrating the distribution of spectral gaps for $\omega_{eg}$ (top panel), $\omega_{fg}$ (middle panel), and $\omega_{fe}$ (bottom panel). Each energy bin has a width of $200 \text{cm}^{-1}$. An increase in the number of counts indicates a corresponding increase in the density of inter-manifold transitions. For a detailed discussion, see Appendix~\ref{app:excitonenergy}.}
  \label{fig:hist}
\end{figure}
The state-specific line-broadening functions are defined as
\begin{align}
    \gamma_{a_{1}^{}}^{} (t) &= \sum_{a_{2}} C_{}(\omega_{a_{1} a_{2}}^{}) \sum_m T_{m a_{1}}^{} T_{m a_{1}}^{} T_{m a_{2}}^{} T_{m a_{2}}^{}
\end{align}
The dissipation function, $C_{}(\omega_{ab})$, is the Fourier transform of the phonon correlation function evaluated at exciton transition frequencies $\omega_{ab}$, expressed as
\begin{align}
   &  C_{}(\omega_{ab}) = \int_0^{\infty} dt \exp(i\omega_{ab} t) \nonumber\\&
    \qquad\times\int \frac{d\omega}{2\pi} J_{\text{ph}}(\omega)\Big[\coth{(\beta \omega/2)} \cos{\omega t} \mp i \sin{\omega t}\Big]
\end{align}
It indicates that the line broadening functions are determined by the phonon spectral function, and exciton gaps. 
The inter-manifold coherence Green's functions contain excitonic resonances. The corresponding values of the inter-manifold dephasing are estimated from the expression of line-broadening by introducing the respective manifold terms. We neglected correlations between the driving source and exciton-phonon dissipation \cite{debnath2013dynamics}.
\section{Note on the exciton energy distribution: one and two-exciton manifold}\label{app:excitonenergy}
As described in Section~\ref{sec:ham}, the exciton system has $14$ one exciton states, denoted $\ket{e_j}$, and $105$ two-exciton states, denoted $\ket{f_k}$. 
The distribution of spectral gaps, the transition energy differences between these manifolds, is evaluated by pairing states from different manifolds. Between the ground and one-exciton states, the number of spectral gaps is $14$, which ranges from $\omega_{e_{1}g_{1}} =E_{e_{1}}$ to $\omega_{e_{14}g_{1}}=E_{e_{14}}$. 
Similarly, the number of spectral gaps between the ground and two-exciton states is $105$, which ranges from $\omega_{f_{1}g_{1}}=E_{f_{1}}$ to $\omega_{f_{105}g_{1}}=E_{f_{14}}$.
In comparison, the spectral gaps between the one and two-exciton manifolds span from $\omega_{f_{1}e_{14}} = 13909.6 \,\text{cm}^{-1}$ to $\omega_{f_{105}e_{1}} =16847.3 \,\text{cm}^{-1}$ contributing a polynomially large number of spectral gaps $14\times 105$, a polynomial large number of spectral gaps. A majority of these gaps are dipole-allowed. They appear in the expressions of the signal via Green's functions. The distribution of these gaps is visualized in Fig.~\ref{fig:hist}.
\section{Two-exciton population distribution}\label{app:signalderiv}
Here, we outline the derivation of the signal presented in Section~\ref{subsec:entexcite}.
The observable is defined using the projection operator onto the desired two-exciton state, averaged over the time-dependent density operator.
Assuming that the perturbative regime of exciton-driving field interaction holds, the density operator can be expanded to fourth-order in incoming field-exciton interaction Hamiltonian to obtain
\begin{widetext}
\begin{align}\label{eqn:popcompact2}
\rho_{ff} (t) &=  (-i)^4 \text{Re}\int^{\infty}_{-\infty} d\tau_4 \int^{\tau_4}_{-\infty} d\tau_3 \int^{\tau_3}_{-\infty} d\tau_2 \int^{\tau_4}_{-\infty} d\tau_1 
\bigex{ P_{+}^{} (t) H_{\text{int},R} (\tau_3) H_{\text{int},R} (\tau_2) H_{\text{int},L} (\tau_4)  H_{\text{int},L}^{} (\tau_1)}
\nonumber\\
&=  (-i)^4 \text{Re}
\int^{\infty}_{-\infty} dt_4 \int^{\infty}_{0} dt_3 \int^{\infty}_{0} dt_2 \int^{\infty}_{0} dt_1 
\ex{P_{+} \mathcal{G}(t_4)d_R  \mathcal{G}(t_3)d_R   \mathcal{G}(t_2)d_L^{\dagger}  \mathcal{G}(t_1)d^{\dagger}_L  }\nonumber\\
&\qquad \ex{ E^{\dagger} (t-t_4)   E^{\dagger} (t-t_4 -t_3)    E (t-t_4 -t_3 -t_2) E (t-t_4-t_3-t_2 -t_1)  }
\end{align}
\end{widetext}
In the expression, the variables $\tau_j$ and $t_j=\tau_{ij}= \tau_i-\tau_j$ denote the exciton-photon interaction times and time-delays, respectively.  The four-point correlation function of the driving field correlate the four-point correlation function of the exciton operators via the time integrals. Expanding the exciton correlation function in the basis of one and two-exciton states, carrying out the averaging over the phonons and inserting the frequency domain variables lead to the expression of the signal.
\begin{widetext}
\begin{align}\label{eqn:popcompact3}
\rho_{ff} (t)  &= 2\, \text{Re}\Big\{ \int \frac{d\omega_4^{}}{2 \pi}\int \frac{d\omega_3^{}}{2 \pi}\int \frac{d\omega_2^{}}{2 \pi}\int \frac{d\omega_1^{}}{2 \pi} 
e^{+ i (\omega_3^{}  +\omega_4^{}  - \omega_2^{} -  \omega_1^{})t }
\ex{E^{\dagger}(\omega_4) E^{\dagger}(\omega_3) E(\omega_2) E(\omega_1) }\nn\\
& \Big\{d_{fe'}^{}d_{e'g}^{} d_{fe}^{*}d_{eg}^{*} \times \mathcal{G}_{ff}^{}(-\omega_4^{}-\omega_3^{}+\omega_2^{}+\omega_1^{} )
  \mathcal{G}_{fe'}^{}(-\omega_3^{}+\omega_2^{}+\omega_1^{})
  \mathcal{G}_{fg}^{}(\omega_2^{}+\omega_1^{})
 \mathcal{G}_{eg}^{}(+\omega_1^{} ) \nn\\
 &+ d_{fe''}^{}d_{fe''}^{*} d_{eg}^{}d_{eg}^{*}\times \mathcal{G}_{ff}^{}(-\omega_4^{}+\omega_2^{}-\omega_3^{}+\omega_1^{})
  \mathcal{G}_{fe''}^{}(+\omega_2^{}-\omega_3^{}+\omega_1^{})
  \mathcal{G}_{e''e'',ee}^{}(-\omega_3^{}+\omega_1^{})
 \mathcal{G}_{eg}^{}(+\omega_1^{} )  \nn\\
 &+d_{fe'}^{}d_{fe}^{*} d_{e'g}^{}d_{eg}^{*} \times \mathcal{G}_{ff}^{}(-\omega_4^{}+\omega_2^{}-\omega_3^{}+\omega_1^{})
  \mathcal{G}_{fe'}^{}(+\omega_2^{}-\omega_3^{}+\omega_1^{}) \mathcal{G}_{ee'}^{}(-\omega_3^{}+\omega_1^{})
 \mathcal{G}_{eg}^{}(+\omega_1^{} ) \nn\\
  &+ d_{fe''}^{*}d_{fe''}^{} d_{eg}^{}d_{eg}^{*} \times \mathcal{G}_{ff}^{}(+\omega_2^{}-\omega_4^{}-\omega_3^{}+\omega_1^{})
  \mathcal{G}_{e''f}^{}(-\omega_4^{}-\omega_3^{}+\omega_1^{})
  \mathcal{G}_{e''e'',ee}^{}(-\omega_3^{}+\omega_1^{})
 \mathcal{G}_{eg}^{}(+\omega_1^{} )  \nn\\
 &+d_{fe}^{*}d_{fe'}^{} d_{e'g}^{}d_{eg}^{*} \times \mathcal{G}_{ff}^{}(+\omega_2^{}-\omega_4^{}-\omega_3^{}+\omega_1^{})
  \mathcal{G}_{ef}^{}(-\omega_4^{}-\omega_3^{}+\omega_1^{}) \mathcal{G}_{e'e}^{}(-\omega_3^{}+\omega_1^{})
 \mathcal{G}_{eg}^{}(+\omega_1^{} ) \Big\}\Big\}
\end{align}\label{eqn:appb3}
\end{widetext}
The expression contains three types of exciton Green's functions that describes inter-manifold coherence, intra-manifold coherence and transport. 
The first term represents the two-exciton coherence-mediated pathways, the second and fourth terms represents one-exciton transport-mediated pathways and the third and fifth terms represents the one-exciton coherence-mediated pathways.
We also note that the Green's functions contains frequencies of the driving field in their arguments. 
The expression can be simplified by plugging in the expressions of the exciton Green's functions. It yields
\begin{widetext}
\begin{align}\label{eqn:popcompact4}
\rho_{ff} (t) 
& \propto 2 \mathrm{Re}\Big\{ \int \frac{d\omega_4^{}}{2 \pi}\int \frac{d\omega_3^{}}{2 \pi}\int \frac{d\omega_2^{}}{2 \pi}\int \frac{d\omega_1^{}}{2 \pi} 
e^{+ i (\omega_3^{}  +\omega_4^{}  - \omega_2^{} -  \omega_1^{})t }
\ex{E^{\dagger}(\omega_4)E^{\dagger}(\omega_3)E(\omega_2)E(\omega_1)}\nn\\
&
\Big\{\frac{d_{fe'}^{}}{-\omega_4^{}-\omega_3^{}+\omega_2^{}+\omega_1^{} - z_{f f}^{}}
\frac{d_{e'g}^{}}{-\omega_3^{}+\omega_2^{}+\omega_1^{} - z_{f e'}^{}}
\frac{d_{fe}^{*}}{+ \omega_2^{}+\omega_1^{}  - z_{fg}^{}}
\frac{d_{eg}^{*}}{\omega_1^{} - z_{eg}^{} }  \nn\\
&+
\frac{d_{fe''}^{}}{-\omega_4^{}+\omega_2^{}- \omega_3^{}+\omega_1^{} - z_{f f}^{}}
\frac{d_{fe''}^{*}}{\omega_2^{}- \omega_3^{}+\omega_1^{} - z_{fe^{''}}^{}}\nn\\
&  d_{e'g}^{}\left( \delta_{e e'} \sum_p \chi^{\text{R}}_{e'' p} D_{pp}^{-1}\frac{1}{- \omega_3^{}+\omega_1^{}- z_p^{}} \chi^{\text{L}}_{p e'} + (1 - \delta_{e e'}) \delta_{e e''} \frac{1}{- \omega_3^{}+\omega_1^{} - z_{e e'}^{}} \right)
\frac{d_{eg}^{*}}{\omega_1^{} - z_{eg}^{}}\nn\\
& +
\frac{d_{fe''}^{*}}{+\omega_2^{}-\omega_4^{}- \omega_3^{}+\omega_1^{} - z_{f f}}
\frac{d_{fe''}^{}}{-\omega_4^{}- \omega_3^{}+\omega_1^{} - z_{fe''}}\nn\\
& d_{e'g}^{}\left( \delta_{e e'} \sum_p \chi^{\text{R}}_{e'' p} D_{pp}^{-1}\frac{1}{- \omega_3^{}+\omega_1^{}- z_p^{}} \chi^{\text{L}}_{p e'} + (1 - \delta_{e e'}) \delta_{e e''} \frac{1}{- \omega_3^{}+\omega_1^{} - z_{e e'}^{}} \right)
\frac{d_{eg}^{*}}{\omega_1^{} - z_{eg}^{}} \Big\}\Big\}
\end{align}\label{eqn:pop1}
\end{widetext}
A compact version of this expression, using the abbreviation $I(\omega_j;  z_{ab}^{}) = (\omega_j+z_{ab}^{})^{-1}$ is presented in Eq.~\ref{eqn:pop1compact}. 
In the final step, tedious yet straightforward frequency integrations yields the expression below
\begin{widetext}\label{eqn:popcompact5}
\begin{align}
\rho_{ff} (t)  
&\propto 2 \text{Re}\Big[\exp{(z_{ff}^{} t)}\Big\{
d_{fe'}^{}d_{e'g}^{}d_{fe}^{}d_{eg}^{*} \times 
\big\langle 
E^{\dagger}(z_{fe'}^{}-z_{ff}^{})
E^{\dagger}(z_{fg}^{}-z_{fe'}^{})
E(-z_{eg}^{}+z_{fg}^{})
E(z_{eg}^{})\big\rangle  \nonumber\\
&+ d_{fe''}^{}d_{fe''}^{*} d_{e'g}^{}d_{eg}^{*} \times
\big\langle 
E^{\dagger}( +z_{fe''}^{}- z_{ff}^{})
E(-z_{p}^{}+z_{fe''}^{})
\sum_p \chi^{R}_{e'' p} D_{pp}^{-1}E^{\dagger}( z_{eg}^{}- z_{p}^{})   \chi^{L}_{p e} 
E(z_{eg}^{})\big\rangle \nonumber\\
&+d_{fe'}^{} d_{fe'}^{*} d_{e'g}^{} d_{eg}^{*}  \times
\big\langle 
E^{\dagger}(+ z_{fe'}^{}- z_{ff}^{})
E( -z_{ee'}^{}+ z_{fe'}^{})
E^{\dagger}( z_{eg}^{}- z_{ee'}^{})  
E( z_{eg}^{})\big\rangle \nonumber\\
&+d_{fe''}^{} d_{fe''}^{*} d_{e'g}^{} d_{eg}^{*}  \times 
\big\langle 
E^{\dagger}(-z_{e''f}^{}+ z_{ff}^{})
E(z_{p}^{}-z_{e''f}^{}) 
\sum_p \chi^{R}_{e'' p} D_{pp}^{-1}E^{\dagger}( z_{e'g}^{}- z_{p}^{})   \chi^{L}_{p e} 
E(z_{eg}^{})\big\rangle \nonumber\\
&+d_{fe''}^{}d_{fe''}^{*} d_{e'g}^{} d_{eg}^{*}\times 
\big\langle 
E^{\dagger}( -z_{ef}^{}+ z_{ff}^{})
E( +z_{ee'}^{}- z_{ef}^{}) 
E^{\dagger}( z_{eg}^{}- z_{ee'}^{}) 
E( z_{eg}^{})\big\rangle  \Big\}\Big]
\end{align} \label{eqn:appb4}
\end{widetext}
This expression is valid for any general correlated photonic source and can be used in numerical simulation.
\section{Entangled photon correlation function}\label{app:entphotoncorr}
Here, we examine the time-frequency properties of entangled photon pairs by plotting the square of the joint spectral amplitude function, using a parameter regime similar to the one described in Section~\ref{sec:results}.
\begin{figure}[ht]
 \includegraphics[width=.48\textwidth]{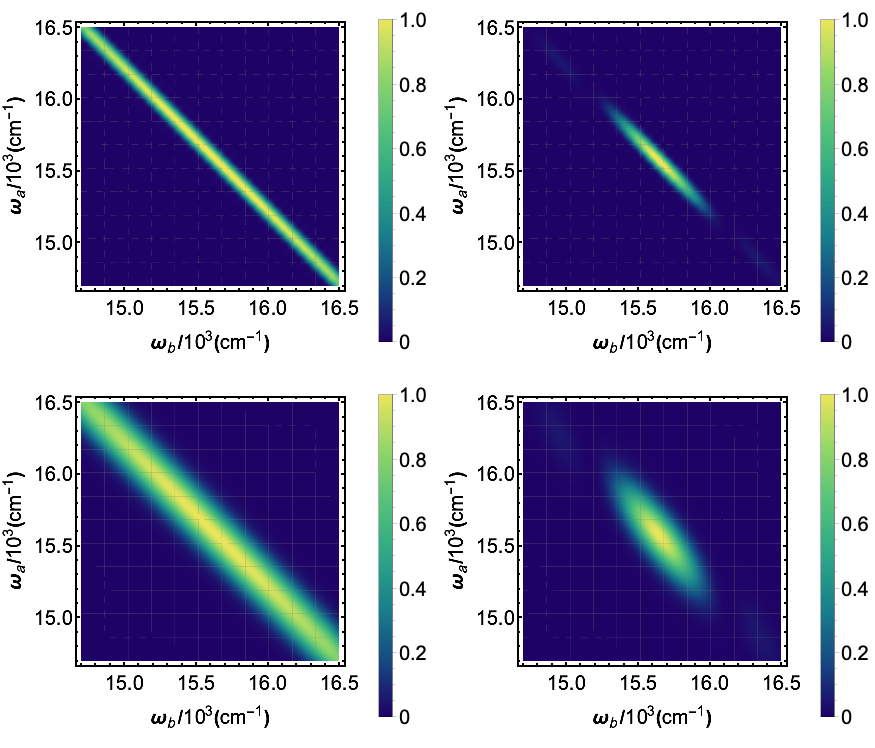}
\caption{The squared joint spectral amplitude (JSA) of the entangled photon pairs, illustrating its dependence on the entanglement time parameter ($\tilde{T}_{\text{ent}}$) and the temporal width of the SPDC pump ($\tau_{0}$). These parameters dictate the frequency correlations or anti-correlations between the signal and idler photons.}  
\label{fig:entcorr81}
\end{figure}
The two-photon wavefunction assumes the functional form:
$\ket{\Psi_{\text{ent}}} \propto \frac{1}{\sqrt{\mathcal{N}}} \iint d\omega_1 d\omega_2 A_p(\omega_1,\omega_2) \text{sinc}\left[ (L/2\pi) \Delta k(\omega_1, \omega_2) \right] \ket{\omega_1, \omega_2}$,
where $\Delta k(\omega_1, \omega_2) = k_1(\omega_1) + k_2(\omega_2) - k_p(\omega_1 + \omega_2)$ is the longitudinal wavevector mismatch, and $\mathcal{N}$ is the state normalization constant. A typical derivation involves expanding the dispersion relation to first order around the central carrier frequencies, i.e., $k(\omega) = k^{(0)} + (\omega - \omega^{(0)})u^{-1}$, where $u$ is the group velocities of the signal-idler pair within the crystal. Under degenerate conditions, we have the entanglement time defined as $\tilde{T}_{\text{ent}} \equiv |u_1^{-1} - u_2^{-1}|L$ \cite{schlawin2018entangled, keller1997theory, schlawin2013suppression}.
The plotted joint spectral amplitude function, as discussed earlier in \ref{subsec:entexcite}, is characterized by diagonal and anti-diagonal widths $\Delta\omega_+$ and $\Delta\omega_-$, respectively. In the limit of a monochromatic pump, one may obtain a stricter bound on the arrival time difference $t_-$ as $|\tau_1 - \tau_2| \leq \tilde{T}_{\text{ent}}$.
\section{Two-photon counting signal}\label{app:gatingderiv}
Here, we present the intermediate steps in the derivation of the photon coincidence counting signal, following the original steps outlined in \cite{dorfman2016time, dorfman2012nonlinear}. Eq~\ref{eq:sigpccbare} already presents the final expression. The bare two-photon counting signal is expressed as:
\begin{align}\label{eqn:gatingN1N2}
& \ex{N_{}(t_1',\tau_1) N_{}(t_2',\tau_2)} =\int_{-\infty}^{t_1'}dt_1\int_{-\infty}^{t_1'+\tau_1}dt_3\nn\\
&\times\int_{-\infty}^{t_2'}dt_2\int_{-\infty}^{t_2'+\tau_2}dt_4
\ex{ V_R^{\dagger}(t_4)V_R^{\dagger}(t_3)V_L(t_1)V_L(t_2)}\nn\\
&\times\sum_{s_1^{}, s_{1'}}\sum_{s_2^{}, s_{2'}} 
\ex{E_{s_{1'} R}(t_4)E_{s_{2'} R}(t_3)
E_{s_1 R}^{\dagger}(t_2'+\tau_2)E_{s_2 R}^{\dagger}(t_1'+\tau_1)\nn\\
&\times E_{s_2 L}(t_1')E_{s_1 L}(t_2')
E_{s_{2'} L}^{\dagger}(t_1)E_{s_{1'} L}^{\dagger}(t_2)}
\end{align}
Here, the field correlation is averaged over the state of the detector modes, assumed to be in vacuum. The exciton correlation function is averaged over the final state after two exciton transport.
We assume that the leading order contributions to the photon counting signal arise from expansion of the density operator to the fourth order in emitted photon field. Expanding the latter into the basis of free-field and following the standard algebraic manipulations we obtain
\begin{align}\label{eqn:nnVVVV}
& \ex{N_{}(t_1',\tau_1) N_{}(t_2',\tau_2)} =|\mathcal{D}(\bar{\omega}_1)\mathcal{D}(\bar{\omega}_2)|^2\\
&\times\langle V_R^{\dagger}(t_2'+\tau_2)V_R^{\dagger}(t_1'+\tau_1)V_L(t_1')V_L(t_2')\rangle.
\end{align}
Here four-point exciton correlation function weighted by the density of detector modes evaluated at emitted frequencies. The exciton correlation function can be expanded further to yield the expressions in ~\ref{eqn:pccpath12}. The Liouville space pathways are depicted in the diagram in Fig.~\ref{fig:diagram}.

\bibliography{main}

\end{document}